\documentclass[preprint, 5p, twocolumn, authoryear, numafflabel]{elsarticle}

\usepackage{amssymb}
\usepackage{amsmath}
\usepackage{hyperref}
\usepackage{adjustbox}
\usepackage{color}
\usepackage{multirow}
\biboptions{compress}
\usepackage{longtable}
\usepackage{multicol}
\usepackage{booktabs}
\usepackage{threeparttablex}

\usepackage{placeins}
\usepackage{enumitem}

\usepackage{graphicx}
\usepackage{subcaption}
\usepackage{caption}
\usepackage{dblfloatfix} 
\usepackage{placeins}
\usepackage{float}
\usepackage{lipsum}
\usepackage{comment}
\usepackage[usenames,dvipsnames]{xcolor}
\usepackage{lineno}

\usepackage{caption}
\usepackage{xcolor}
\definecolor{myblue}{RGB}{0,0,255} 

\usepackage{tikz}
\usetikzlibrary{shapes.geometric, arrows.meta, positioning, fit, backgrounds}

\usepackage{hyperref}
\hypersetup{
    colorlinks=true,
    linkcolor=blue,      
    citecolor=blue,      
    urlcolor=blue        
}

\journal{Physics of the Dark Universe}

\begin{document}

\begin{frontmatter}

\title{\textbf{Unidentified Sources in the 4FGL-DR4 \textit{Fermi}-LAT Catalog as Potential Dark Matter Satellites}} 

\author[inst1,inst2]{Fernando Valenciano}
\ead{fernando.valenciano@iac.es}

\affiliation[inst1]{organization={Instituto de Astrofísica de Canarias (IAC)},
            addressline={C/ Vía Láctea, s/n}, 
            city={La Laguna},
            postcode={E-38205}, 
            state={Canary Islands},
            country={Spain}}

\affiliation[inst2]{organization={Departamento de Astrofísica, Universidad de La Laguna},
            state={Santa Cruz de Tenerife},
            country={Spain}}

\author[inst3,inst4]{Miguel A. Sánchez-Conde}

\affiliation[inst3]{organization={Instituto de Física Teórica (UAM-CSIC)},
            addressline={C/ Nicolas Cabrera 13-15}, 
            postcode={E-28049}, 
            state={Madrid},
            country={Spain}}

\affiliation[inst4]{organization={Departamento de Física Teórica M-15},
            addressline={Universidad Autónoma de Madrid}, 
            postcode={E-28049}, 
            state={Madrid},
            country={Spain}}

\author[inst1,inst2]{Jorge Martin Camalich}

\begin{abstract}
The ultimate nature of dark matter (DM) remains unknown, standing as one of the most pressing questions in modern physics. In this work we focus on indirect DM detection via gamma-ray emission from annihilating weakly interacting massive particles (WIMPs). Within the $\Lambda$CDM cosmological framework, the hierarchical assembly of structure predicts a rich population of DM halos and subhalos. If WIMPs compose DM, annihilation inside these subhalos would generate gamma rays that could be detectable by instruments such as the \textit{Fermi}-LAT. Since 2008, \textit{Fermi}-LAT has cataloged several thousands of gamma-ray sources, with a third lacking clear astrophysical association at present. This opens the exciting possibility that some of these unassociated sources (unIDs) may actually be dark satellites, i.e., Galactic subhalos not massive enough to retain gas or stars at all and thus with no  counterparts at other wavelengths. Starting from the 14-year \textit{Fermi}-LAT catalog (4FGL-DR4), we conduct a systematic search for dark satellite candidates among its 2428 unIDs. We first design and apply a series of filters based on expected properties of dark satellites to reduce our sample, and then consider two different scenarios depending on the level of restriction of these rejection criteria. Through an exhaustive spectral \textit{Fermi}-LAT data analysis, we find that none of the remaining unIDs shows a statistically significant preference for a DM interpretation. Then, in the absence of a clear hint, we place conservative constraints on the $\langle\sigma v\rangle-m_{\chi}$ DM parameter space, which are a factor of 2--3 above the thermal relic cross section value at the lowest WIMP masses in a conservative scenario, and are able to rule out thermally-produced WIMPs below $\sim$50 GeV in the $b\bar{b}$ channel when assuming the sensitivity reach of the method. Overall, and despite the lack of a clear signature of DM annihilations among the pool of unIDs, our results emphasize the distinctive value of unID-based searches: they yield constraints that complement those derived from classical DM targets, such as dwarf spheroidal galaxies, and probe a unique, cosmologically motivated population of dark satellites that is otherwise inaccessible to observations.
\end{abstract}




\end{frontmatter}

\section{Introduction}
\label{intro}
\noindent
Over the past century, a wealth of compelling evidence has supported the existence of a non-baryonic form of dark matter (DM) that accounts for roughly 25\% of the total energy and matter content of the Universe \citep{2018RvMP...90d5002B, 2020A&A...641A...6P}. Historically, this evidence includes rotation curves of spiral galaxies \citep{1970ApJ...159..379R, 1983SciAm.248f..96R, 1989A&A...223...47B, 1985ApJ...295..305V}, velocity dispersions in galaxy clusters \citep{1932BAN.....6..249O, 1937ApJ....86..217Z}, gravitational lensing \citep{1990ApJ...350...23B, 1998MNRAS.295..587M, 2006ApJ...648L.109C}, and measurements of the anisotropies of the Cosmic Microwave Background (CMB) \citep{2020A&A...641A...6P}. 
Despite considerable efforts, the ultimate nature of this DM remains elusive, constituting a primary objective in contemporary physics.
The absence of a viable DM candidate within the Standard Model (SM) has prompted searches for physics beyond it, with extensions to high energies in which new particles arise, some of which could be natural candidates for DM (see \cite{2005PhR...405..279B, 2015JPhG...43c3001G, Hook:2019sB}). 
Among these, \textit{Weakly Interacting Massive Particles} (WIMPs; \citealp{2010Natur.468..389B, 2018RPPh...81f6201R}) are particularly well-motivated: a stable DM particle with a thermally averaged annihilation cross section of $\langle \sigma v \rangle \simeq 2.2\times10^{-26}\text{ cm}^3\text{ s}^{-1}$ is predicted to freeze out of equilibrium in the early Universe leading to a constant relic abundance equal to the cosmological abundance of DM; this appealing coincidence is known as the \textit{WIMP miracle} \citep{2012PhRvD..86b3506S, Hooper:20190U}.

Indirect searches for DM signatures in astrophysical environments are a powerful tool to explore the DM properties. If a WIMP candidate self-annihilates into SM particles, with an annihilation cross section close to the thermal value, this would yield a gamma-ray signature 
that could be bright enough to be detectable by gamma-ray telescopes \citep{2015JETP..121.1104C}, such as the \textit{Fermi}-LAT \citep{2009ApJ...697.1071A},
MAGIC \citep{2016APh....72...76A}, H.E.S.S. \citep{2006A&A...457..899A}, VERITAS \citep{ 2006APh....25..391H} or the upcoming CTAO \citep{2019scta.book.....C}.
To test such a scenario, various astrophysical targets have already been examined, including the Milky Way Galactic center \citep{2009arXiv0910.2998G, 2015PhRvD..91f3003C, 2017ApJ...840...43A, 2021PhRvD.103l3005D}, 
dwarf spheroidal galaxies (dSphs, e.g. \cite{2016JCAP...02..039M, 2024PhRvD.109f3024M}), the extragalactic gamma-ray background \citep{2015ApJ...800L..27A, 2015JCAP...09..008F},
galaxy clusters \citep{2015ApJ...812..159A, 2023PhRvD.107h3030D}, dwarf irregular galaxies \citep{2021PhRvD.104h3026G}, stellar streams \citep{2025JCAP...09..003F} and dark subhalos \citep{2015JCAP...12..035B, 2019JCAP...07..020C, 2025PhRvD.111b3033C}.

In the standard $\Lambda$CDM cosmology, numerical simulations reveal a hierarchical structure formation in a bottom-up scenario: the smaller structures form and virialize first in the Universe and then, by accretion and merging of these, large and massive structures, or halos, arise \citep{2012AnP...524..507F, 2019Galax...7...81Z}. 
Consequently, individual DM halos contain a large number of smaller subhalos \citep{2008ApJ...679.1260M, 2008MNRAS.391.1685S}.
The minimum halo mass depends on the DM candidate, typical expected values ranging from $10^{-3}M_{\odot}$ down to $10^{-11}M_{\odot}$ \citep{2009NJPh...11j5027B} for WIMPs. However, current state-of-the-art numerical simulations can only resolve structures above $\sim 10^5M_{\odot}$ \citep{2008Natur.454..735D,  2008MNRAS.391.1685S, 2024MNRAS.532.1814G}. 
Besides, in the case of subhalos, it is currently unclear whether the smallest $\Lambda$CDM structures will survive the strong tidal forces they are subject to within their hosts, see, e.g., \cite{2003ApJ...584..541H, 2018MNRAS.475.4066V, 2020MNRAS.491.4591E, 2023MNRAS.518...93A}.
The most massive subhalos will retain baryons and lead to the formation of dwarf galaxies; however, the vast majority of lower-mass subhalos are expected to remain completely ``dark'' -- lacking significant baryonic content or visible counterparts. The threshold for this transition is governed by a complex interplay of physical processes, including reionization history, gas cooling, and assembly bias. Consequently, the minimum host halo mass required to support viable galaxy formation is estimated to lie within the range $10^6-10^9 M_{\odot}$ \citep{2015MNRAS.448.2941S, 2016MNRAS.456...85S, 2018MNRAS.473.2060J, 2020ApJ...893...48N, 2022ApJ...940....8M, 2025ApJ...983L..23N}. We will follow previous works and adopt a threshold subhalo mass of $10^7M_{\odot}$ for these dark satellites \citep{2026PDU....5202262P}.

These \textit{dark satellites} -- smaller, baryon-devoid analogues of dSphs, lacking multi-wavelength (MW) counterparts but potentially lying closer to Earth (at typical distances of a few kpc; e.g.~\cite{2024MNRAS.530.2496A}) -- may appear as unidentified sources (unIDs) in gamma rays if made of WIMPs, thus offering promising targets for DM searches.

Since its launch in 2008, the Large Area Telescope on board the NASA \textit{Fermi Gamma-ray Space Telescope} (\textit{Fermi}-LAT) has been surveying the gamma-ray sky in the energy range from 30 MeV to more than 1 TeV. 
The most recent data release source catalog from the \textit{Fermi}-LAT instrument, the 4FGL-DR4 \citep{2023arXiv230712546B, 2022ApJS..260...53A},
covering 14 years (2008--2022), contains 7195 non-transient sources, of which 2428 are unIDs, i.e. sources without a clear association to a known gamma-ray source, or to any other MW counterpart. This relative fraction of unIDs has been a consistent feature across all catalogs since the initial 1-year release, in which 630 out of 1451 were unassociated \citep{2010ApJS..188..405A}. Despite significant efforts to elucidate the nature of these objects, many with successful results, new unIDs have emerged over time due to increased exposure time. Interestingly, the overall fraction of unIDs appears to be constant over time, with one third of unIDs in each catalog. The increasing number of \textit{Fermi}-LAT unIDs opens the exciting possibility that some of these sources might be the result of WIMP self-annihilation or decay in Galactic dark satellites.

The main goal of this work is to pinpoint those unIDs in the 4FGL-DR4 catalog that may be compatible with being dark satellites, according to their spectral and spatial properties, by applying a series of filters built from our current knowledge of expected DM emissions. We also use this information to provide upper limits to the DM annihilation
cross section, which are competitive and complementary to those obtained by means of other targets and methods. 
We note that, while the majority of the previous DM constraints were obtained by assuming the absence of a gamma-ray signal, in this study we consider the possibility that the observed gamma-ray signal in unIDs may have originated from DM annihilation. Consequently, the adopted methodology will follow previous works done in this context~\citep{2019JCAP...07..020C, 2019JCAP...11..045C, 2021PDU....3200845C, 2022PhRvD.105h3006C, 2020Galax...8....5C}, which relies on a comparison between the number of 4FGL-DR4 unIDs that are compatible with being dark satellites, and that predicted by N-body simulations of Milky Way-sized halos \citep{2008Natur.454..735D, 2024MNRAS.530.2496A}.
The methodology also provides an accurate statistical treatment to set strong limits on the properties of the DM particle, combining results from both observations and N-body cosmological simulations. 
In this sense, this work represents an update of the work done in \cite{2019JCAP...07..020C, 2019JCAP...11..045C} for the 3FGL \citep{2015ApJS..218...23A}, 2FHL \citep{2016ApJS..222....5A} and 3FHL \citep{2017ApJS..232...18A} \textit{Fermi}-LAT catalogs.
Here we adopt not only the most recent 4FGL-DR4 catalog, which includes six more years of data than the original 4FGL \citep{2020ApJS..247...33A}, 
but also implement a series of significant improvements in different aspects of the analysis, including a more refined modeling of the DM spectra; the adoption of a larger number of DM annihilation channels; a more sophisticated search of MW counterparts of unIDs; improved LAT sensitivity estimates to DM signals; and a more robust statistical interpretation of the results.

The structure of the work is outlined as follows. In Section \ref{sec1} the expected gamma-ray flux from DM annihilation in dark satellites is described. In Section \ref{sec2} we discuss and establish the rejection criteria used to find and select potential dark satellites among the pool of  4FGL-DR4 unID sources.  
We consider two different subsamples, based on  different levels of restrictiveness of the criteria. 
In Section \ref{Sec-4} we conduct a dedicated LAT analysis of the best dark satellite candidates, which also includes a rigorous evaluation of the DM hypothesis against an astrophysical one for the origin of the unID emission. In Section \ref{Sec-5} we set new and competitive constraints on the DM parameter space, by comparing the number of unIDs that survive our filters with the number of predicted dark satellites in N-body simulations, conveniently repopulated with low-mass subhalos below the original resolution limits. We conclude in Section \ref{Sec-6}.

\section{Expected dark matter-induced gamma-ray flux}\label{sec1}

\noindent
The differential gamma-ray flux, $\phi_{\gamma}$, induced by WIMP annihilation occurring within an astrophysical target can be computed as \citep{Hooper:20190U, 2020Galax...8...25R},
\begin{equation}\label{gamma_flux}
\begin{split}
\phi_{\gamma}(E_{\gamma},\Delta\Omega)_{\rm ann} &= \frac{1}{4\pi}\frac{\langle\sigma v\rangle}{2m_{\chi}^2}\sum_iB_i\left(\frac{dN_{\gamma}}{dE_{\gamma}}\right)_i \\
&\quad \times \int_{\Delta\Omega}\int_{\rm l.o.s.}\rho_{\chi}^2(l,\Omega)\,dl\,d\Omega.
\end{split}
\end{equation}
where $\langle\sigma v\rangle$ is the averaged DM annihilation cross section, $m_{\chi}$ the WIMP mass, $(dN_{\gamma}/dE_{\gamma})_i$ is the differential gamma-ray spectrum per annihilation for the channel $i$, and $B_i$, the branching ratio, refers to the fraction of annihilations that goes into a particular channel $i$.
The term outside the integral in Eq. (\ref{gamma_flux}) encapsulates the particle physics information, while the integral codifies all relevant astrophysical considerations. The latter term is usually referred to as the astrophysical factor or J-factor 
\citep{1998APh.....9..137B, 2019JCAP...07..020C},
where $\rho_{\chi}$ is the DM density profile (e.g., ~\cite{1996ApJ...462..563N}), and the integral is performed along the line of sight ($l.o.s.$), $\Delta\Omega$ being the solid angle of observation. 

As for $dN_{\gamma}/dE_{\gamma}$, this term is usually computed via Monte Carlo event generators such as 
PYTHIA \citep{2015CoPhC.191..159S}, DarkSUSY \citep{2018JCAP...07..033B} or micrOMEGAs \citep{2024CoPhC.29909133A}. Starting from such low-level event data, more user-friendly packages exist that provide WIMP annihilation (and decay) spectra for a large number of possible channels, e.g., PPPC 4 DM ID 
\citep{2011JCAP...03..051C} and, more recently, CosmiXs \citep{2024JCAP...03..035A}. The latter is the one we will adopt in the present work (see Section \ref{Sec-3.2} for further details).

\section{Identification of dark satellite candidates in the 4FGL-DR4}\label{sec2}

\subsection{Initial sample selection}\label{Sec-3.1}
\vspace{1mm}

\begin{figure*}[t]
    \centering
    \includegraphics[width=0.65\linewidth]{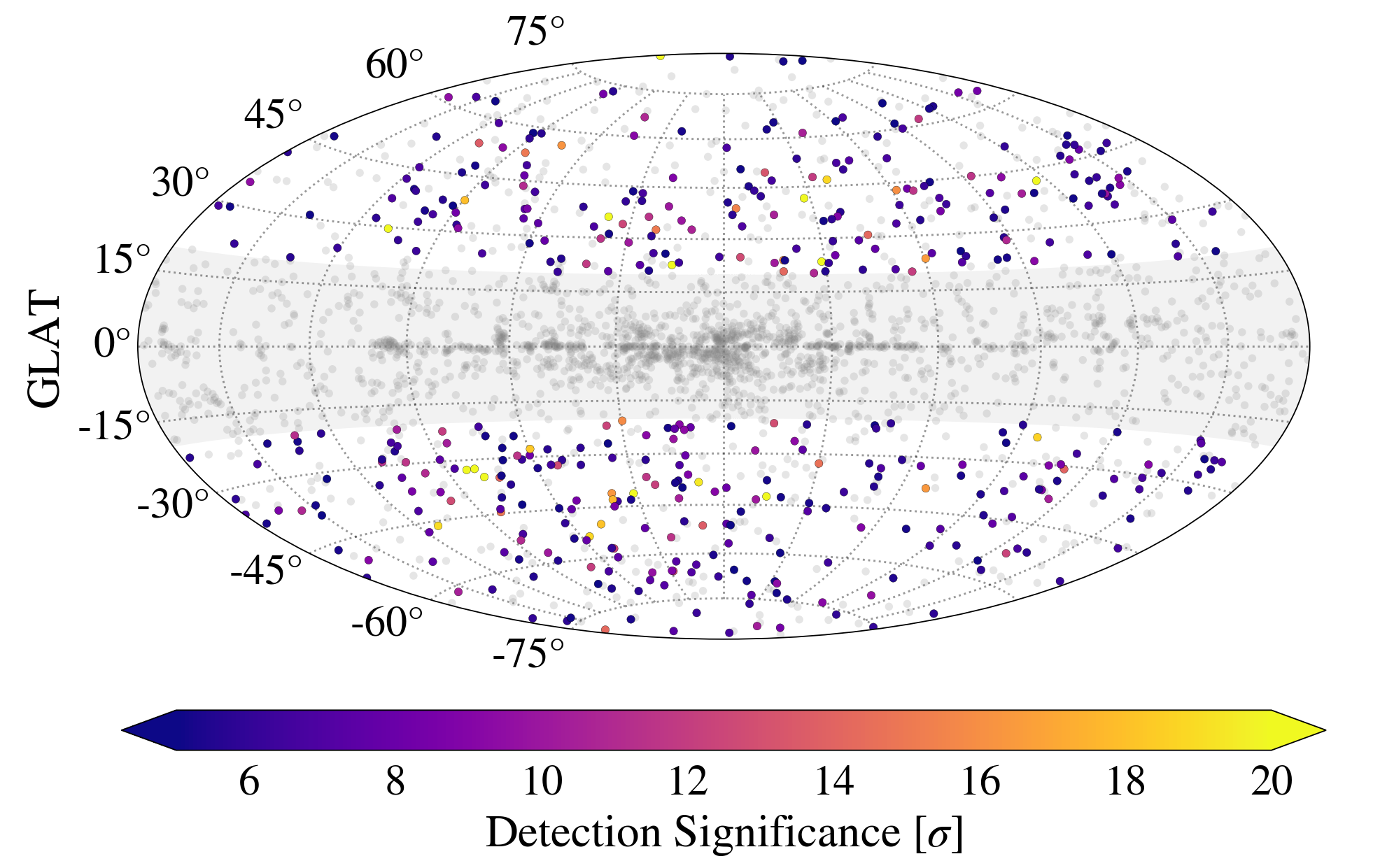}
    \caption{Sky location of 4FGL-DR4 unIDs in Galactic coordinates and Hammer projection. Color indicates the detection significance of each source, as depicted in the color bar. The gray area refers to the latitude cut introduced in our analysis to avoid the Galactic plane region; see Section~\ref{Sec-3.1}. Only sources passing the initial selection criteria -- latitude, variability and detection significance -- are highlighted in color, leaving a final sample of 445 unIDs.} 
    \label{plot-hammer-unids}
\end{figure*}

\begin{figure*}[t]
    \centering
    \includegraphics[width=\linewidth]{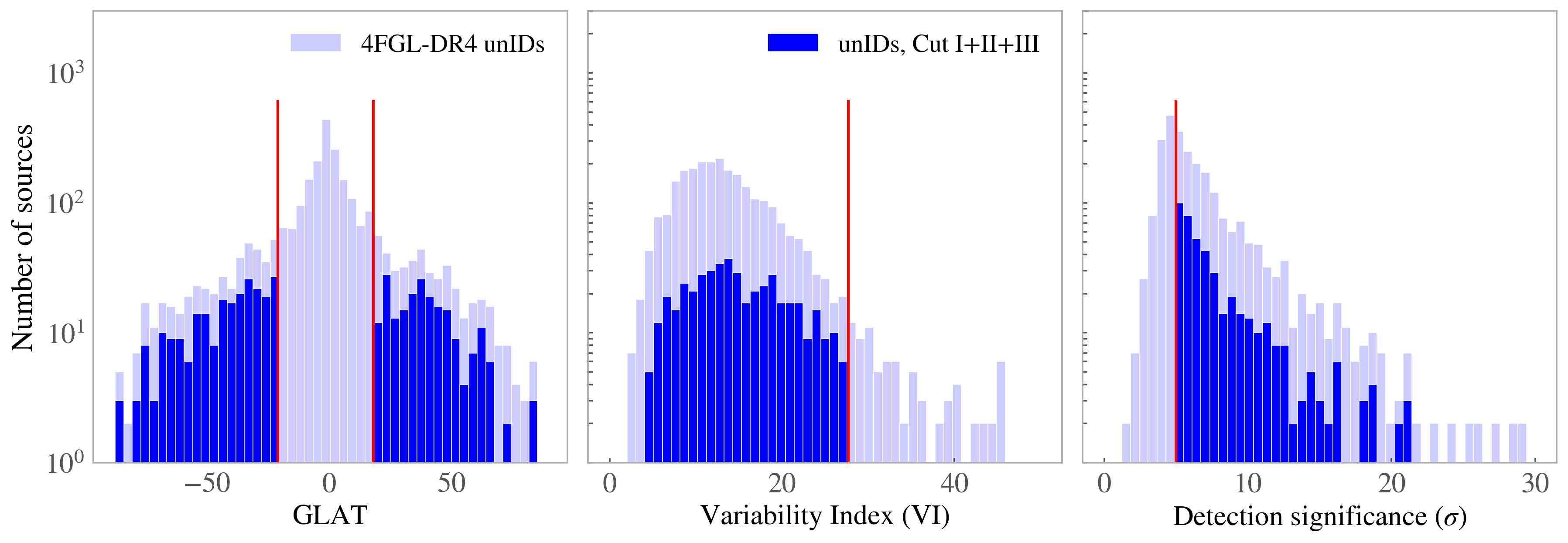}
    \caption{Selection criteria used in Section \textcolor{blue}{\ref{Sec-3.1}} to build a first sample of dark satellite candidates among the initial 2428 4FGL-DR4 unIDs. In each panel, the shaded blue represents the whole sample of 4FGL-DR4 unIDs, while those surviving our cuts are depicted in dark blue. \textit{Left:} Galactic latitude distribution (GLAT); sources between the red lines, representing $|b|=20^{\circ}$, are excluded from our search (`Cut I'). \textit{Middle:} Variability index distribution; the red line represents the $\text{VI}=27.69$ threshold, above which sources are considered variable in the catalog at 99\% C.L. and are thus excluded from our study (`Cut II'). \textit{Right:} Detection significance, with a threshold at the $5\sigma$ value adopted for our sample (`Cut III').}
    \label{plot-cuts-unids}
\end{figure*}

\noindent
The 4FGL-DR4 \citep{2023arXiv230712546B} is an incremental version (Data Release 4) of the fourth \textit{Fermi}-LAT catalog (4FGL; \citealp{2020ApJS..247...33A}); it covers the first 14 years of the mission, spanning energies between 50 MeV and 1 TeV, and comprises 7195 gamma-ray sources, of which 2428 correspond to unIDs. Fig. \textcolor{blue}{\ref{plot-hammer-unids}} shows the sky location of the overall 4FGL-DR4 unID population. Of the remaining two-thirds, the vast majority of the sources are classified as Active Galactic Nuclei (AGNs) \citep{2022ApJS..263...24A}, including BL Lacertae blazars (BL Lacs) and flat-spectrum radio quasars (FSRQs), Seyfert galaxies and radio galaxies, with a total of 4278 AGN-like sources. The remainder, constituting a smaller population, typically consists of Galactic pulsars and millisecond pulsars (MSPs), accounting for 320 sources. As our first task, we create a list of potential dark satellites among the number of unIDs present in the 4FGL-DR4. The chosen criteria are based on the expected properties of dark satellites compared to the ones of astrophysical sources:
\vspace{2mm}

\noindent
\textbf{i. Latitude:} From numerical cosmological simulations we expect dark satellites to be isotropically distributed over the Galactic halo, thus with no preference for any particular latitude. In contrast, Galactic sources such as pulsars or supernova remnants are concentrated along the Galactic plane, resulting in a non-homogeneous distribution with a pronounced peak at around $b=0^{\circ}$ (as can be observed in Fig. \ref{plot-hammer-unids} and in the left panel of Fig. \ref{plot-cuts-unids}). Moreover, pulsars may also contaminate our sample due to the similarity of their spectra with that of DM annihilations (see, e.g., \citealp{2008JCAP...07..013B, 2013MNRAS.436.2461M}). Also, most low-latitude unIDs show soft, curved spectra, and a strong concentration towards the plane, possibly linked to mismodeled diffuse emission \citep{2026ApJ..1006..188A}. Consequently, for the purpose of this work, we decide to exclude those unIDs with $|b|\leq20^{\circ}$. It should be noted that there may be low-latitude dark satellites among the rejected sources. Yet, this conservative cut will help simplify our DM search by removing mainly unIDs with an astrophysical origin. Out of the initial 2428 unIDs, more than half are rejected after applying such cut, resulting in a sample of 858 sources. We refer to this filter as `Cut I'.
\vspace{2mm}

\noindent
\textbf{ii. Variability:} Another predicted feature of the DM annihilation flux is that it should be steady.
The \textit{Fermi}-LAT catalogs provide a parameter known as variability index (VI), which indicates the probability of a source being variable based on a $\chi^2$ statistic. To be conservative in building our sample, we remove sources marked as variable at 99$\%$ C.L. (<1\% chance of being steady) in the 4FGL-DR4, i.e., those sources exhibiting a value of $\text{VI}>27.69$ in the catalog. From the previous 858 sources, 780 meet our variability criterion (`Cut II'), see the middle panel of Fig. \ref{plot-cuts-unids}.
\vspace{2mm}

\noindent
\textbf{iii. Detection significance:} Finally, we  remove those sources that have not reached the 5$\sigma$ significance threshold required to have a robust detection (see the right panel of Fig. \ref{plot-cuts-unids}). Again, while there might be fainter dark satellites among the unIDs rejected this way, the data analysis results that could be derived from such weak sources would be of low quality and insufficient to conclude anything robust on their nature, needless to say to make a substantive claim of any kind. Moreover, some of such weak sources in the catalog could be statistical artifacts due to mismodeling of the Galactic diffuse emission. In addition, the statistical test we will later apply to test a DM hypothesis is not accurate for faint sources. Thus, the previous sample of 780 high-latitude and steady unIDs is now reduced to a final sample of 445 (`Cut III').

\subsection{unIDs spectral scrutiny: $\beta$-plot and subsamples}\label{Sec-3.2}
\label{sec-3.2}
\begin{figure*}[!hb]
    \centering
    \includegraphics[width=0.49\linewidth]{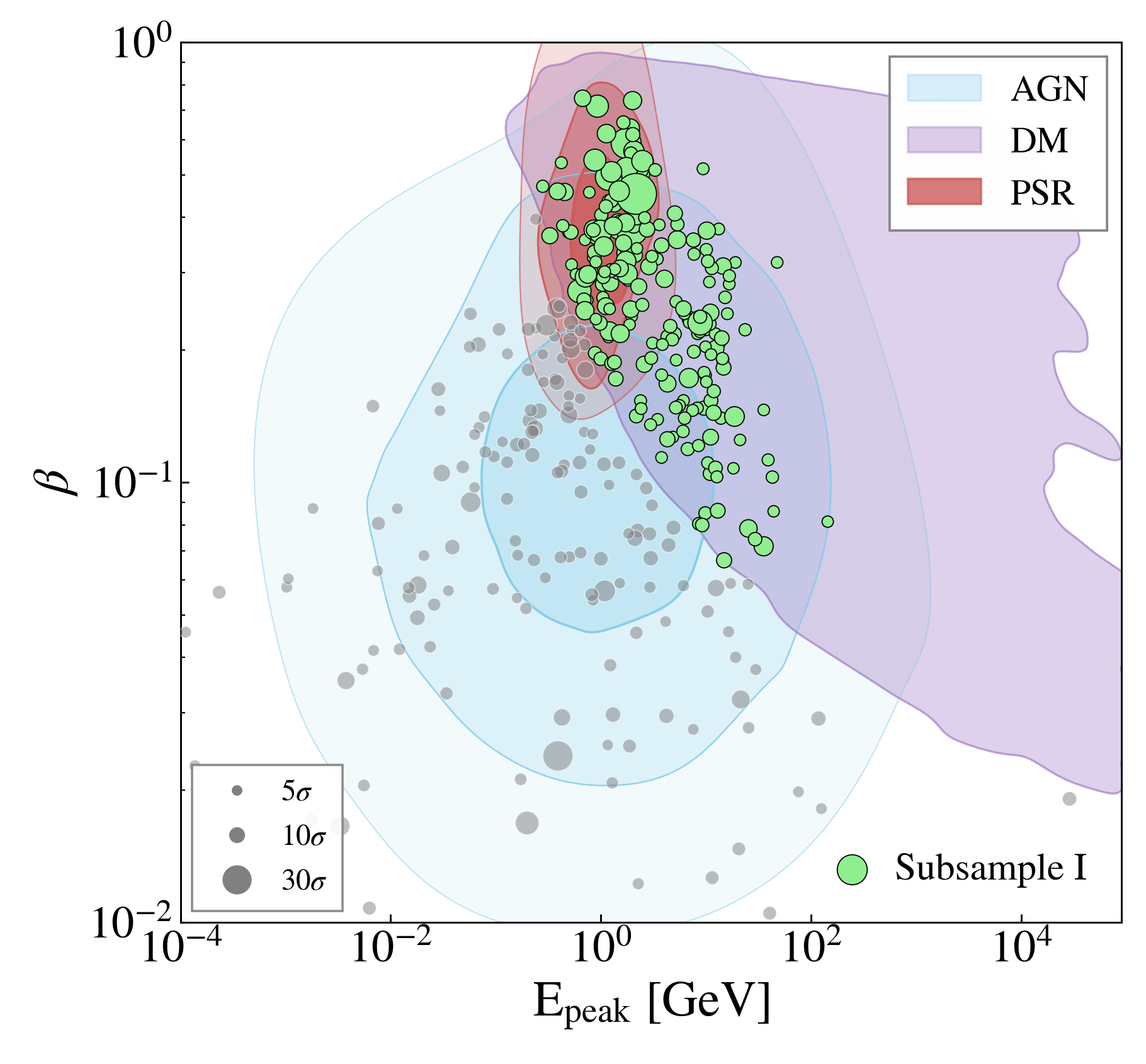} 
    \includegraphics[width=0.49\linewidth]{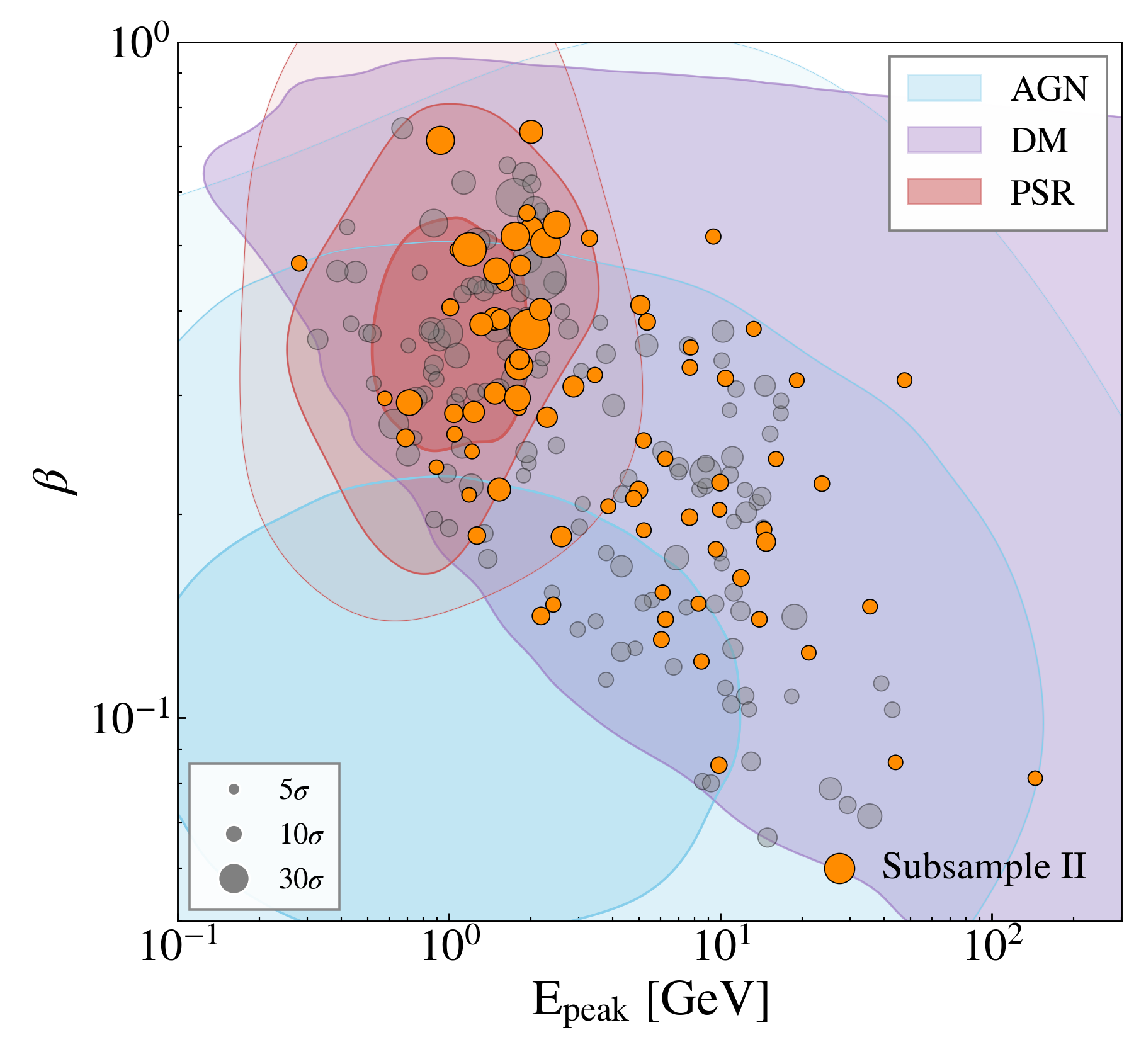}
    \caption{$E_{\rm peak}-\beta$ parameter space (so-called $\beta$-plot) built from the LogParabola parametrization of Eq. (\ref{LogParabola}) for AGNs (blue), pulsars (red) -- illustrated by filled contours, marking their one, two and three-sigma outline -- and DM (purple). The latter region is obtained by fitting a DM annihilation spectrum for all possible two-channel combinations, for DM masses from $5$~GeV to 100~TeV. 
    The two unID subsamples defined in our study are also shown in these plots; 
    see Section \ref{Sec-3.2} for details. \textit{Left}: In green, the 220 unIDs belonging to Subsample I. The rest of the 445 unIDs that survived our initial selection cuts in Section~\ref{Sec-3.1} are depicted in gray. \textit{Right}: A zoomed-in region of the $\beta$-plot is shown, with the unIDs of Subsample II highlighted in orange. Now, gray points refer to those unIDs in Subsample I that were excluded after our multi-wavelength analysis.}
    \label{plot-subsamples}
\end{figure*}

\noindent
Spectral confusion between astrophysical sources, such as pulsars, and DM annihilation signals has been a persistent challenge in indirect DM searches, especially for the $b\bar{b}$ channel, whose spectrum for certain masses is indistinguishable from that of a pulsar, e.g., \cite{2008JCAP...07..013B,2013MNRAS.436.2461M}. To address this confusion we will make use of a discrimination technique that relies on the so-called $\beta$-plot \citep{2019JCAP...11..045C, 2023MNRAS.520.1348G}. The methodology consists of studying the unIDs in a derived parameter space defined by the peak energy ($E_{\rm peak}$) and curvature ($\beta$) parameters in the LogParabola (LP)  spectral parametrization, 
\begin{equation}\label{LogParabola}
    \frac{dN}{dE}=K\left(\frac{E}{E_0}\right)^{-\Gamma-\beta\log(E/E_0)}
\end{equation}
where $K$ is the normalization, $\Gamma$ the photon spectral index, $\beta$ the spectral curvature index, and $E_0$ the energy scale. $E_{\rm peak}$ is the energy at which the energy flux reaches its maximum, given by:
\begin{equation}
    E_{\rm peak}=E_0\times\text{e}^{\frac{2-\Gamma}{2\beta}}
\end{equation}
In the $E_{\rm peak}-\beta$ parameter space, a clear distinction can be made between pulsars and AGNs, which is particularly effective for brighter sources. 
However, when considering fainter sources, this distinction becomes less clear, resulting in an overlap between them \citep{2023MNRAS.520.1348G}.
This can be seen in Fig. \ref{plot-subsamples}, where the pulsar population (red) exhibits higher $\beta$ values and a more concentrated $E_{\rm peak}$ distribution compared to that of AGNs (blue), 
with lower curvature and $E_{\rm peak}$ reaching higher values\footnote{We note that the 4FGL-DR4 includes a Gaussian prior on the $\beta$ values, with mean set to 0.1 and width to 0.3. This is done so as to have a width narrow enough to avoid the very large $\beta$ values $\sim 1$ that were present in previous catalogs, which was thought to be indicative of a fake $\beta$ value in fainter sources \citep{2023arXiv230712546B}.}.

The technique shows its full potential when DM annihilation signal predictions are included.
As first done in \cite{2019JCAP...11..045C}, we include DM into the $E_{\rm peak}-\beta$ parameter space by first generating multiple DM spectra and then fitting them to a LP spectral function. We use CosmiXs \citep{2024JCAP...03..035A} to generate the spectrum for each possible DM channel and allowed DM mass from 5 GeV to 100 TeV, which we note represents an important refinement with respect to \cite{2019JCAP...11..045C}, which used \cite{2011JCAP...03..051C} instead.
Furthermore, while in previous studies only pure annihilation channels were considered, our work explores a more general scenario that allows for a two-channel combination:
\begin{equation}
    \frac{dN}{dE}=B_i\left(\frac{dN}{dE}\right)_{c_1}+(1-B_i)\left(\frac{dN}{dE}\right)_{c_2}
\end{equation}
where $c_1$ and $c_2$ are the two possible channels. We combine them by adopting 11 different branching ratio values, ranging from 0 to 1 in steps of 0.1, and for 100 DM mass values ranging between 5 GeV and 100 TeV. Being agnostic to the specific DM particle physics model, we treat all these synthetic DM points as belonging to a ``DM region'' in the $\beta$ plot, with the goal of having three classes in it, namely AGNs, pulsars, and predicted DM spectra, as shown in Fig. \ref{plot-subsamples}. Interestingly, although there is significant overlap among classes in such parameter space, some regions are unique to DM.

\noindent
Fig. \ref{plot-subsamples} (left) illustrates the location of the ($E_{\rm peak},\beta$) values of the 445 unIDs that were derived from the 4FGL-DR4 after having applied the initial filtering procedure described in Section~\ref{Sec-3.1}. As will be discussed in Section \ref{Sec-5}, the number of unIDs that survive our filters plays a crucial role in the limits we can set on the WIMP parameter space: the fewer the candidates, the stronger the constraints we can impose. Because of that, in the following we consider two different scenarios based on the adopted level of restrictiveness.
\\

\noindent
\textit{Subsample I: all unIDs compatible with DM}
\vspace{1mm}

\noindent
The first and most straightforward filter we can apply is to retain only those unIDs whose ($E_{\rm peak},\beta$) values fall within the boundaries of the DM region in the $\beta$-plot. It should be noted that our DM \textit{cloud}, unlike those associated with pulsars and AGNs, represents a complete population, unbiased by statistics or instrumental performance. Therefore, this initial filter includes those 4FGL-DR4 unIDs that, based on their spectrum, are theoretically compatible with DM. This allows us to remove approximately half of the sources, leaving us with 220 unIDs.
\vspace{3mm}

\noindent
\textit{Subsample II: MW emission}
\vspace{1mm}

\noindent
If the gamma-ray signal of our unIDs is indeed produced by DM annihilations in nearby dark satellites 
--- i.e., light subhalos with DM masses below $10^7M_{\odot}$ and devoid of baryonic contents --- then we do not expect any MW emission other than gamma rays.
We note this assumption may not hold exactly, since most annihilation channels inject $e^\pm$ that could produce secondary IC/synchrotron emission (e.g., \citealp{2006A&A...455...21C}). However, dark satellites are expected to lack the gas and stars needed to sustain a significant magnetic field or radiation field, so injected leptons should escape largely uncooled without efficiently radiating. Overall, this secondary DM flux is largely uncertain, and typically expected to be subdominant \citep{2004PhRvD..70b3512B, 2016JCAP...11..021L}. We thus consider a MW counterpart within the error region to remain a robust indicator of astrophysical origin.
This is not the case for standard gamma-ray astrophysical emitters. For instance, AGNs are expected to peak at radio and gamma-ray energy ranges. Then, in order to identify the most suitable dark satellite candidates we remove from our sample those that exhibit other wavelength emission.
To do this, we first exclude those unIDs that in the 4FGL-DR4 dataset have a secondary classification (CLASS2) assigned. In the 4FGL-DR4 catalog, CLASS2 does not indicate an alternative association but rather flags a low-confidence counterpart \citep{2022ApJS..260...53A, 2023arXiv230712546B}; sources retaining the unID label therefore still lack a high-confidence identification, but the presence of such a hint makes them less suitable as dark satellite candidates, thus we discard them.
Then, we revisit current literature searching for any possible counterparts at other wavelengths. In particular, we use:
\begin{enumerate}[label=(\roman*)]
    \item Recent catalogs created with the Very Large Array Sky Survey (VLASS) to pinpoint radio counterpart candidates among the \textit{Fermi}-LAT unIDs \citep{2021ApJ...914...42B, 2023ApJ...943...51B}. Although no associations are claimed, we prefer to stay conservative and exclude these potential associations from our list. 
    \item Studies of X-ray counterparts via machine-learning classification with the Chandra X-ray Observatory (CXO) \citep{2024ApJ...971..180Y}, probabilistic cross-matching with eROSITA \citep{2024A&A...684A.208M} and neural-network classification with Swift-XRT \citep{2021ApJ...923...75K}, or via optical spectroscopic follow-up of Swift-XRT counterparts \citep{2024MNRAS.530.4626U}. As in (i), no robust associations are claimed in these works --- they assign classification probabilities or spectroscopic types to the candidate counterpart itself, but its link to the \textit{Fermi}-LAT source still rests on positional coincidence within the X-ray refined error region. We therefore conservatively exclude these potential counterparts from our list.
    \item The \textcolor{blue}{\href{https://simbad.cds.unistra.fr/simbad/}{SIMBAD}} repository, where we manually search for additional emission in the area of each unID. In particular, we reject the unID if any kind of additional emission is found within a 5 arcmin radius of the cataloged source position in the 4FGL-DR4. We note that this does not mean that the unID can be associated with the found emission, for which a more detailed analysis is mandatory. For the purposes of our analysis, though, rejecting the unID in such cases represents a conservative choice, not only in our search for potential dark satellites but also when setting DM constraints later in Section~\textcolor{blue}{\ref{Sec-5}}.
\end{enumerate}

\begin{table}[!t]
    \centering
    \caption{Summary of rejected unIDs in Subsample II after our MW counterpart search. The ``Order'' column gives the sequence of criteria: (1) low-confidence secondary classification in the CLASS2 field of 4FGL-DR4 \citep{2022ApJS..260...53A, 2023arXiv230712546B}; (2) radio counterpart candidates from VLASS-based catalogs \citep{2021ApJ...914...42B, 2023ApJ...943...51B}; (3)~X-ray counterpart candidates from machine-learning or probabilistic classifications with Chandra \citep{2024ApJ...971..180Y}, eROSITA \citep{2024A&A...684A.208M}, and Swift-XRT \citep{2021ApJ...923...75K}, or from spectroscopic follow-up of Swift-XRT counterparts \citep{2024MNRAS.530.4626U}; (4)~additional emission within a 5 arcmin radius of the 4FGL-DR4 position in SIMBAD. Counts give the number of unIDs rejected at each step among those not already excluded by earlier criteria. See \textcolor{blue}{\ref{app-B}} for the per-source breakdown.}
    
    \begin{tabular*}{\columnwidth}{@{\extracolsep{\fill}}clc}
    \hline
    \hline
    Order & Reason & unIDs rejected \\
    \hline 
    1 & CLASS2 & 25 \\
    2 & Radio  & 79 \\
    3 & X-ray  & 29 \\
    4 & SIMBAD & 10 \\
    \hline
    \end{tabular*}
    \label{table-cuts}
\end{table}

\noindent
With the above MW study, summarized in Table \textcolor{blue}{\ref{table-cuts}}, we are able to exclude 143 sources. 
Each of these was tested against every criterion of our MW search, rather than being removed at the first match, so that the full record of rejection reasons per source could be retained (see \textcolor{blue}{\ref{app-B}}). The remaining 77 unIDs are highlighted in the right panel of Fig. \textcolor{blue}{\ref{plot-subsamples}} in the $\beta$-plot, and represent our Subsample II.

\begin{figure*}[t]
    \centering
  \includegraphics[width=0.85\linewidth, trim=0 2cm 0 3cm, clip]{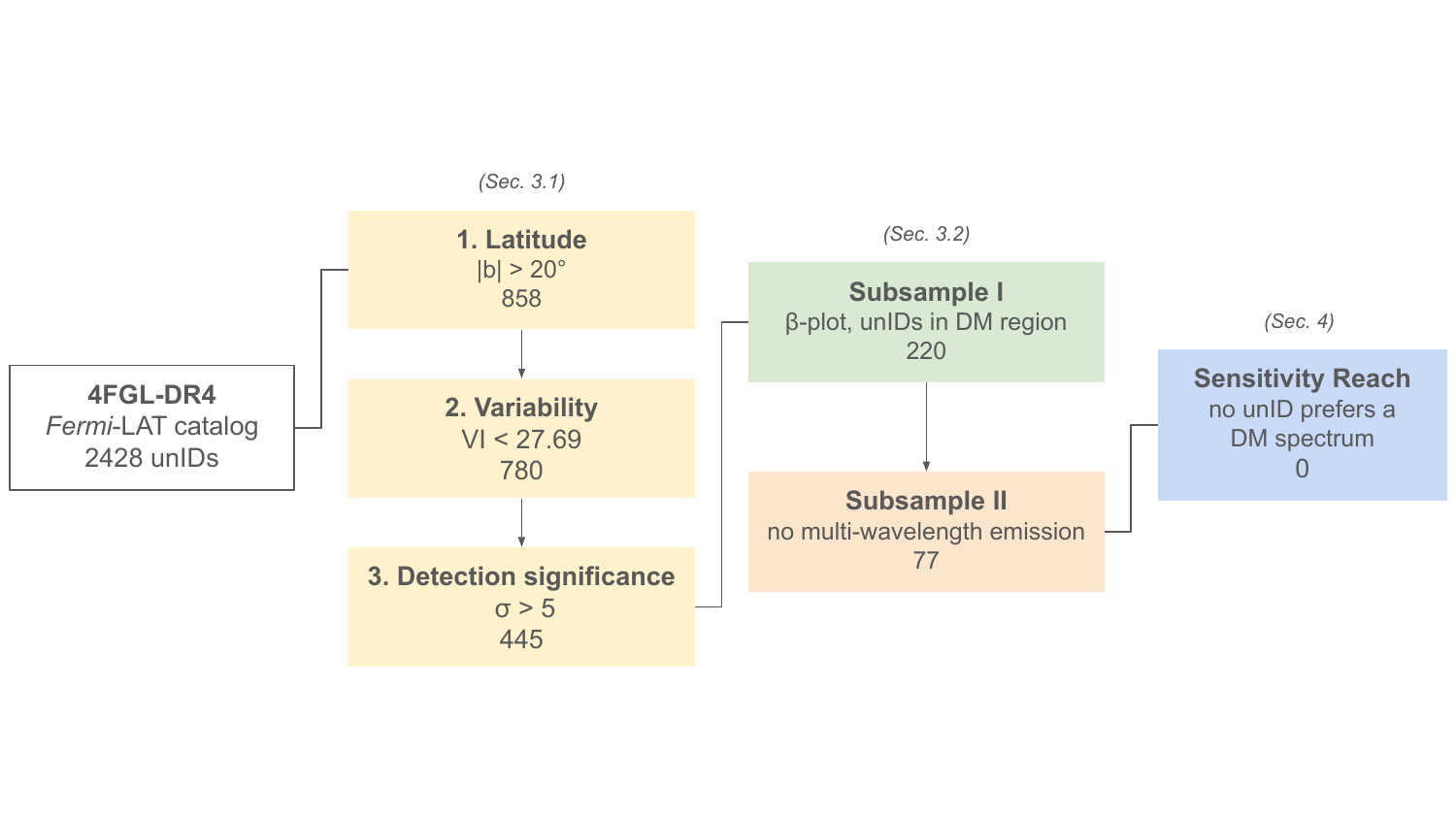}
    \caption{Flowchart of the unIDs filtering procedure applied to search for dark satellite candidates among the unIDs in the 4FGL-DR4 \textit{Fermi}-LAT catalog. The first three rejection criteria or {\it cuts} are introduced in Section \textcolor{blue}{\ref{Sec-3.1}}, while the definition of Subsamples I and II can be found in Section \textcolor{blue}{\ref{Sec-3.2}}. The sensitivity reach (SR) scenario is described in Section \textcolor{blue}{\ref{Sec-4}}, and corresponds to the case of finding no unIDs compatible with DM properties after an exhaustive spectral analysis.}
    \label{fig-flowchart}
\end{figure*}

The flowchart of Fig. \textcolor{blue}{\ref{fig-flowchart}} provides a general overview of the unID selection performed throughout this section, with the initial filters detailed in Section \textcolor{blue}{\ref{Sec-3.1}} and our definition of Subsamples I and II in Section \textcolor{blue}{\ref{Sec-3.2}}. The final scenario, defined as sensitivity reach, will be presented in the following section.

\section{\textit{Fermi}-LAT spectral analysis of dark satellite candidates}\label{Sec-4}

\noindent
In this section, we perform a full dedicated spectral analysis of our Subsample II, composed of 77 unIDs,
with the aim of confirming or refuting their potential DM origin.  

\subsection{Technical setup and likelihood analysis}\label{Sec-4.1}

\noindent
The gamma-ray data obtained by \textit{Fermi}-LAT, after passing specific selection criteria, are made publicly available via the LAT Data Server\footnote{\url{https://fermi.gsfc.nasa.gov/cgi-bin/ssc/LAT/LATDataQuery.cgi}}
of the FSSC, typically within 24 hours. These data can be analyzed with \textit{Fermipy 1.2.2}\footnote{\url{https://fermipy.readthedocs.io/en/latest/}}, a Python software package that provides a high-level interface for LAT analysis \citep{2017ICRC...35..824W}, whose core is based on the \textit{Fermi ScienceTools 2.2.0}\footnote{\url{https://fermi.gsfc.nasa.gov/ssc/data/analysis/software/}}.

The setup starts by defining a configuration file containing the essential parameters of the
analysis, including data selection (time and energy intervals), region of interest (ROI) and model specifications. 

Pass 8 events \citep{2013arXiv1303.3514A} are used with the instrumental response functions (IRFs) specified by \texttt{P8R3\_SOURCE\_V3}. We use FRONT+BACK events, with the Galactic diffuse and isotropic templates given by \texttt{gll\_iem\_v07.fits} and \texttt{iso\_P8R3\_SOURCE\_V3\_v1.txt} \citep{2018arXiv181011394B}, respectively. Table~\ref{tab_config} summarizes the main parameters in the setup of our LAT analysis.

\begin{table}[!b]
    \small
    \centering
    \caption{Summary of the \textit{Fermi}-LAT data analysis setup used for the unIDs in Subsample II; see Section~\ref{Sec-4} for details.}
    \begin{tabular}{ll}
    \hline \hline
        Temporal range (MET) & (239557417, 733662371)\\
        Temporal range (ISO 8601) & (2008-08-04, 2024-04-01) \\
         Energy range (MeV) & (50, 1000000) \\
         IRFs & \texttt{P8R3\_SOURCE\_V3}\\
         Event type & FRONT+BACK\\
         ROI radius & $15^{\circ}$\\ 
         Angular bin size & 0.1$^{\circ}$\\
         Zenith angle cut (ZMAX) & $90^{\circ}$\\
         Bins per energy decade & 8\\
         Galactic diffuse model & \texttt{gll\_iem\_v07.fits}\\
         Isotropic diffuse model & \texttt{iso\_P8R3\_SOURCE\_V3\_v1.txt}\\
         \hline 
    \end{tabular}
    \label{tab_config}
\end{table}

The first step involves running \texttt{fermipy setup()}, which calls the \textit{Fermi ScienceTools} responsible for preparing the data and the model to initiate the analysis.
An optimization of the background model parameters follows, which uses \texttt{fermipy optimize()} to determine the best-fit spectral parameters for all components of the model. This is typically achieved by performing a maximum likelihood fit to obtain the optimal values of the model parameters. The likelihood $\mathcal{L}$ represents the probability of obtaining the data given an input model. It is generally defined as:
\begin{equation}
    \mathcal{L}=e^{-N_{\rm exp}}\prod_i\frac{m_i^{n_i}}{n_i!}
\end{equation}
where $m_i$ is the expected number of counts in bin $i$, $n_i$ the detected counts and $N_{\rm exp}$ the counts predicted by the model, i.e. the sum of $m_i$. These quantities are related to the $\chi^2$ statistical function. To define the significance of a source detection, the Test Statistic (TS) parameter is used
\begin{equation}\label{eq_ts}
    \text{TS}=-2\times\log\left[\frac{\mathcal{L}(H_0)}{\mathcal{L}(H_1)}\right]
\end{equation}
where $\mathcal{L}(H_0)$ and $\mathcal{L}(H_1)$ are the likelihoods of the null hypothesis (no source) and the alternative hypothesis (existing source), respectively. The detection threshold is usually placed at TS$~=25$, which corresponds to a detection significance of about $5\sigma$ (TS$~\sim \sigma^2$) \citep{10.1214/aoms/1177732360, 10.1214/aoms/1177728725}. 
The likelihood analysis is performed by maximizing the $\mathcal{L}$ function, whose value provides the parameters that best fit the data. In practice, this is done in our analysis by: (i) freeing the normalization of the sources with \texttt{fermipy free$\_$sources()} within 3$^{\circ}$ from the center of the ROI, (ii) searching for new sources within the ROI with \texttt{fermipy find$\_$sources()}, (iii) performing the fit with \texttt{fermipy fit()}, and (iv) executing \texttt{fermipy optimize()} to optimize the model parameters. 
To inspect the goodness of the fit, residual maps showing the difference between data and model are created with \texttt{fermipy residmap()}, together with TS detection maps generated via \texttt{fermipy tsmap()}.

The spectrum of the source is generated by \texttt{fermipy sed()}. Since the best-fit spectral parameters depend on the adopted model, we adopt the most widely used spectral parametrizations: PowerLaw (PL), LogParabola (LP) and PowerLaw with SuperExponential-CutOff (PLSEC)\footnote{\url{https://fermi.gsfc.nasa.gov/ssc/data/analysis/scitools/source_models.html}.}. 
The publicly available version of \textit{Fermipy} allows for a specific spectral function for DM annihilations, based on the \texttt{DMFitFunction} which is derived from previous DM annihilation predictions made by \cite{2008JCAP...11..003J}, which differ from those presented in PPPC4 \citep{2011JCAP...03..051C} and CosmiXs \citep{2024JCAP...03..035A}, mainly for heavy DM masses. 
    We have updated the \textit{Fermipy} pipeline to optionally use CosmiXs gamma-ray predictions.\footnote{This update will be made available to the general public with the next official \textit{Fermipy 1.4.4}.}
    For a pure annihilation channel it follows
    \begin{equation}
        \frac{dN}{dE}=K\times F_{m_{\chi}, i}
    \end{equation}
    where $K=\langle \sigma v\rangle\times J$ is the normalization and $F_{m_{\chi},i}$ the photon spectrum per annihilation for channel $i$ and DM mass $m_{\chi}$.

\subsection{Statistical preference for DM over  astrophysical spectral models}\label{Sec-4.2}

\noindent
Following the outlined procedure, a spectral analysis of the 77 unIDs that were found to be compatible with DM is performed (Subsample II, see Section \textcolor{blue}{\ref{Sec-3.2}}). Through this analysis, our objective is to assess whether any of these sources exhibits a statistically significant preference for a DM interpretation over standard astrophysical models. 
To compare the spectral fits, we employ the corrected Akaike Information Criterion (AIC$_c$) \citep{1974ITAC...19..716A, 2007MNRAS.377L..74L}, defined as
\begin{equation}\label{eq-deltats}
    \text{AIC}_c=-2\ln\mathcal{L}_{\rm max}+2k +\frac{2k(k+1)}{N-k-1}
\end{equation}
where $\mathcal{L}_{\rm max}$ is the maximum likelihood achievable by the model, $k$ the number of parameters in the model ($k=2$ for \texttt{DMFit} and PL, $k=3$ for LP and $k=4$ for PLSEC) and $N$ the number of data bins entering the binned likelihood. The best model is the one which minimizes the AIC$_c$
\footnote{The standard AIC is defined as $\text{AIC}=-2\ln\mathcal{L}_{\rm max}+2k$, and is only strictly valid asymptotically, i.e.\ when the number of data points $N$ is much larger than the number of free parameters $k$ ($N/k\gtrsim40$). The additional term $2k(k+1)/(N-k-1)$ in Eq.~(\ref{eq-deltats}) corrects for this small-sample bias, more strongly penalizing models with more parameters when $N$ is not much larger than $k$. In our case, this is relevant given the limited number of spectral data points available per source in our sample.}. 
Then, for each source, we define 
\begin{equation}
\label{delta_aic}
    \Delta \text{AIC}_c=\text{AIC}_{c, \text{DM}}-\text{AIC}_{c,\text{astro}}
\end{equation} which can be expressed as a Bayesian probability of preferring a DM interpretation by the Akaike weights \citep{2007MNRAS.377L..74L}:
\begin{equation}
\label{omegaDM}
    \omega_{\rm DM}=\frac{1}{1+e^{\Delta\text{AIC}_c/2}}
\end{equation}
For each source, we test the three astrophysical models (PL, LP and PLSEC) and adopt the one that minimizes the AIC$_c$ value, even when it differs from the spectral type listed in the 4FGL-DR4 catalog\footnote{The 4FGL-DR4 catalog is based on 14 years of \textit{Fermi}-LAT data (2008 August--2022 August). In this work we analyze almost 16 years, thus differences in the preferred spectral model with respect to the catalog may arise from the different integration times.}.
\begin{figure*}[!h]
    \centering
    \includegraphics[width=\linewidth]{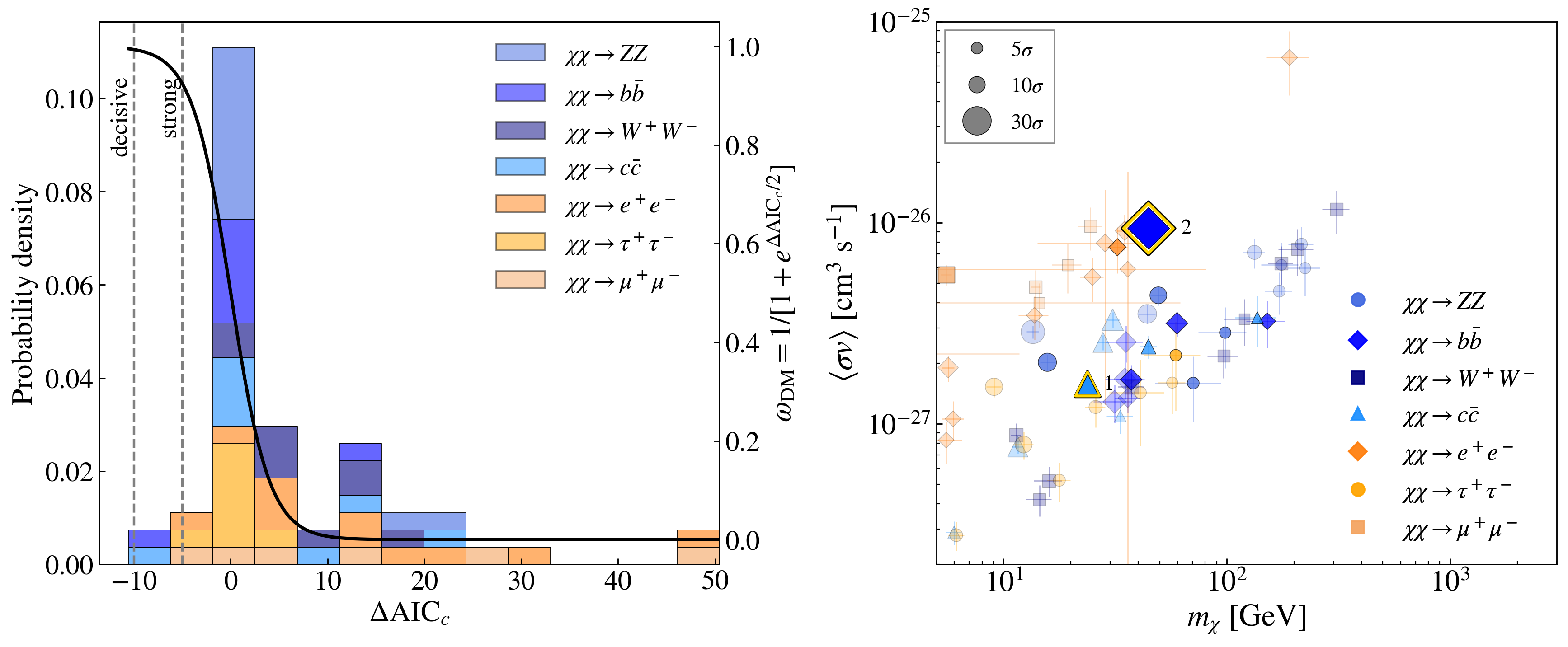}
    \caption{Results of the unID spectral analysis to address the statistical preference for DM over standard astrophysical models. \textit{Left}: $\Delta$AIC$_c$ (best DM channel vs astrophysical model) probability density distribution (see Eq.~\textcolor{blue}{\ref{delta_aic}}) and the corresponding $\omega_{\rm DM}$ values, i.e. probability of preferring a DM fit (Eq.~\textcolor{blue}{\ref{omegaDM}}), computed for the 77 unIDs of Subsample II that were found to be well-fitted to a DM spectrum without MW counterparts (see Section~\textcolor{blue}{\ref{Sec-3.2}}). Only the best-fit DM channel is shown in each case, with the colors given in the legend.
\textit{Right}: Best-fit $\langle\sigma v\rangle$ and $m_{\chi}$ values for each unID in Subsample II. Marker size scales with the detection significance, $\sigma \sim \sqrt{\rm TS}$, and marker opacity encodes the sign of $\Delta$AIC$_c$: sources with $\Delta$AIC$_c>0$ are rendered with reduced opacity, while those with $\Delta$AIC$_c\leq0$ are rendered with higher opacity, so that the DM-preferred sources stand out visually. The two sources whose $\Delta$AIC$_c$ falls in the nominal (pre-trials) strong-evidence regime ($\Delta{\rm AIC}_c\leq-5$) are additionally highlighted with a black-and-gold halo around their marker and are numbered 1 and 2 (1: 4FGL J2342.4-4739, 2: 4FGL J1225.9+2951); none of the 77 sources falls in the decisive-evidence regime ($\Delta{\rm AIC}_c\leq-10$). Although a few of them show mildly negative $\Delta$AIC$_c$ values (left panel), indicating a modest preference for a DM interpretation, the results in this $\langle\sigma v \rangle-m_{\chi}$ plane show no clustering at all and, thus, do not support the idea of having a single, common DM model capable of explaining the emission of all these unIDs, as would be expected from a genuine DM interpretation.}
    \label{fig:deltaaic}
\end{figure*}

For the case of DM spectra, we perform spectral fits for different pure channels: $e^+e^-$, $\mu^+\mu^-$, $\tau^+\tau^-$, $b\bar{b}$, $c\bar{c}$, $W^+W^-$ and $ZZ$ with DM masses from 5 GeV to 100 TeV.\footnote{No kinematic threshold is imposed on the DM mass, so in the $W^+W^-$ and $ZZ$ channels
$m_\chi$ can lie below $m_W$ or $m_Z$. In this regime, CosmiXs includes annihilation into off-shell gauge
bosons ($VV^*$, $V^*V^* \to 4f$), which its authors consider adequate down to $m_\chi = 5$~GeV (private communication). The spectra are normalized per annihilation, so the fitted
$\langle\sigma v\rangle$ does not include the off-shell suppression. Below $m_V/2$ both bosons are
off-shell, a rate strongly suppressed in realistic models, and the photon spectrum resembles a quark
cascade. Such fits should therefore be read as a description of the spectral shape, not as evidence
for annihilation into gauge bosons.}
For the purpose of this work, we will assume a single channel as potential best-fit for a single source spectrum, although \texttt{DMFitFunction} allows the incorporation of branching ratios different from one.
The left panel of Fig. \textcolor{blue}{\ref{fig:deltaaic}} summarizes the resulting $\Delta$AIC$_c$ values for the best-fit DM channels as well as the corresponding DM preference probability $\omega_{\rm DM}$, while the right panel of the figure shows the best-fit $\langle \sigma v\rangle$ and $m_{\chi}$ values for each of the 77 sources in Subsample II. 
Two sources fall below the threshold for \textit{strong evidence} ($\Delta\text{AIC}_c\leq-5$, \cite{2007MNRAS.377L..74L}), which are listed in \ref{App-E}. For these sources, a dedicated follow-up study will be performed elsewhere.  
Overall, although these few sources exhibit a mild tendency towards a DM fit, the global distribution indicates a clear preference for astrophysical explanations. Moreover, the results in the $\langle  \sigma v\rangle-m_{\chi}$ plane (right panel) show no clustering of the best-fit DM values, providing no indication of a single, common DM model that could simultaneously account for several unIDs' gamma-ray spectra, as would be expected if they actually were dark satellites. 

It should also be noted that the $\Delta$AIC$_c$ values are pre-trials, and do not account for the fact that multiple DM channels are tested per source; any trial correction would further reduce the relative preference for DM. Motivated by this result, in the following we also consider an additional scenario in which none of the unIDs is a dark satellite, hereafter sensitivity reach (SR) scenario. This scenario is the one potentially capable of providing the most stringent constraints using the adopted methodology. We caution the reader, however, that these constraints should be regarded as predictions rather than actual constraints for the time being, until each of the 77 unIDs in Subsample II has been robustly ruled out as a potential dark satellite. 

\section{Constraints on the DM particle properties}\label{Sec-5}

In the absence of an unequivocal DM annihilation signature among the pool of 4FGL-DR4 unIDs, in this section we proceed to set limits on the WIMP properties. We will do so for the different scenarios (i.e., unID subsamples) introduced in this work. First, in Section~\ref{Sec-5.1}, we  describe the adopted methodology and compute the different ingredients that are needed to set the DM limits. Then, in Section~\ref{Sec-5.2}, we obtain these under different setups and compare them with other DM limits in the literature.

\subsection{Procedure to set DM constraints}\label{Sec-5.1}

\noindent
The procedure to set DM constraints starting from our unID work involves a comparison between the predicted number of dark satellites found in simulations that should be detectable with \textit{Fermi}-LAT, and the actual number of remaining unIDs that might be potential dark satellites~\citep{2019JCAP...07..020C, 2019JCAP...11..045C, 2021PDU....3200845C, 2022PhRvD.105h3006C, 2020Galax...8....5C}. The master equation for setting constraints on the WIMP annihilation cross section involves the use of Eq. (\textcolor{blue}{\ref{gamma_flux}}) to relate $\langle\sigma v\rangle$ with $m_{\chi}$ in the following way:
\begin{equation}\label{eq-constraints}
    \langle\sigma v\rangle=\frac{8\pi~F_{\rm min}}{N_{\gamma}~J}\times m_{\chi}^2
\end{equation}
where $F_{\rm min}$ is the (LAT) instrumental sensitivity to DM annihilations, $J$ the J-factor, and $N_{\gamma}$ the expected number of photons per annihilation. In the following, we detail the methodology employed to determine these quantities.
\\

\noindent
\textit{A. \textit{Fermi}-LAT sensitivity to dark satellites, $F_{\rm min}$:}
\vspace{1mm}

\noindent
$F_{\rm min}$ describes the minimum flux that is required in order to achieve a $5\sigma$ detection with \textit{Fermi}-LAT. Previous literature on the subject, e.g. {\cite{2015JCAP...12..035B, 2012A&A...538A..93Z}, often assumed a constant \textit{Fermi}-LAT sensitivity to DM, equal to the sensitivity threshold of the catalog\footnote{In this case, $F_{\min,\text{4FGL-DR4}}\sim1.5\times10^{-12}\text{ erg cm}^{-2}\text{ s}^{-1}$ \citep{2023arXiv230712546B}.}. However, it is important to note that such sensitivity is derived assuming a power-law spectrum, $dN/dE \propto E^{-2}$, typical of astrophysical gamma-ray emitters, yet not accurate enough to represent most DM spectra. Indeed, the minimum flux required for a DM detection depends on the considered DM annihilation channel, DM mass, and sky location. In our work, we follow the work in ~\cite{2019JCAP...07..020C} and compute the actual \textit{Fermi}-LAT sensitivity to WIMP annihilation in dark satellites, by generating all-sky sensitivity maps for each DM annihilation channel and mass. We do so by placing a putative dark satellite at each of the 49152 equal-area pixels of a HEALPix $N_\mathrm{side}=64$ all-sky map (pixel size $\approx 0.92^\circ$) 
and obtaining the integrated flux that is required to obtain TS $=25$ (see Eq.~\ref{eq_ts}) in each case, i.e., a $\sim5\sigma$ detection. We employ the \textit{Sensitivity Tools} of \textit{Fermipy}, with the following inputs: the DM spectral parameters (set by annihilation channel and mass); the TS threshold (here set to 25); a minimum number of photon counts (set to 3); the diffuse and isotropic templates (\texttt{gll\_iem\_v07.fits} and \texttt{iso\_P8R3\_SOURCE\_V3\_v1.txt}, respectively); the preferred energy range (we use 50 MeV -- 1 TeV, as in the catalog); adopted spatial morphology (point-like); and a specific type of HEALPix pixelation (NESTED ordering with $N_\mathrm{side}=64$). This way, we obtain all-sky sensitivity maps for each of the 7 annihilation channels considered in this work and 30 DM masses (210 maps in total).

Once the $|b|\leq20^{\circ}$ region is masked, similar to our `Cut I' in Section \ref{Sec-3.1}, we compute the mean and standard deviation of $F_{\rm min}$ over the remaining regions of the sky. Fig. \textcolor{blue}{\ref{fig-fmin}} illustrates the obtained \textit{Fermi}-LAT sensitivity and 1$\sigma$ uncertainty bands for different DM channels and masses. In general, the harder spectra of leptonic channels yield higher sensitivities over most of the considered WIMP mass range, though some hadronic channels ($W^+W^-$ and $ZZ$) perform better at the high mass end\footnote{We note that Fermipy sensitivity tool models the background with the Galactic diffuse and isotropic templates only, i.e. without including 4FGL-DR4 sources. Thus, the resulting $F_{\rm min}$ values correspond to isolated sources and represent a lower limit on the actual detection thresholds. We expect this effect to be negligible for our sample, as our candidates are high-latitude isolated sources, and their potential DM annihilation fluxes are typically orders of magnitude above $F_{\rm min}$.}. The horizontal dashed line in Fig.~\ref{fig-fmin} shows the nominal 4FGL-DR4 detection threshold, $F_{\rm 4FGL\text{-}DR4}\simeq1.4\times10^{-9}\,\mathrm{ph\,cm^{-2}\,s^{-1}}$ (0.1--100~GeV). 
Note that hadronic channels fall well below this line already for $m_\chi\gtrsim20$--$30$~GeV, becoming around one (two) order(s) of magnitude more sensitive than catalog-level detection at WIMP masses $\sim$100 GeV ($\sim$100 TeV). Compared to the DM sensitivity study performed for the 3FGL catalog setup in \cite{2019JCAP...07..020C}, our extended energy range (up to 1 TeV versus 300 GeV) plus ten additional years of data allow for a sensitivity enhancement of roughly a factor of two, especially for leptonic channels.
Note also that our work makes use of the latest DM annihilation predictions from CosmiXs \citep{2024JCAP...03..035A}, thus further refining current DM sensitivity expectations.
\\

\noindent
\textit{B. Number of photons per annihilation, $N_{\gamma}$:}
\vspace{1mm}

\noindent
To compute $N_{\gamma}$, we perform the following integration:
\begin{equation}
    N_{\gamma}=\int^{E_{\rm max}}
_{E_{\rm th}}\left(\frac{dN}{dE}\right)dE,
\end{equation}
where $(dN/dE)$ is the predicted differential annihilation spectrum obtained using CosmiXs, as described in Section \ref{Sec-3.2}. The upper limit of the integral corresponds to the considered DM mass, $E_{\text{max}}=\min(m_{\chi}, 1\text{ TeV})$, so that photons undetectable by the instrument are not counted, and $E_{\rm th}$ is the adopted energy threshold, set to 50 MeV in the 4FGL-DR4 catalog \citep{2023arXiv230712546B}. Numerical integration and further interpolation are conducted to obtain the total number of gamma-ray photons for any given DM mass value, the latter ranging between 5 GeV and 100 TeV. Fig. \textcolor{blue}{\ref{fig-ngamma}} shows the results of $N_\gamma$ for different annihilation channels considered. Hadronic channels yield large photon counts, owing to the decay of  neutral pions  produced in hadronization. Leptonic channels produce far fewer photons (up to two orders of magnitude less, depending on channel and WIMP mass), as their spectra are dominated by final-state radiation and lack the hadronic cascade; the $\tau^+\tau^-$ is intermediate, since it also decays hadronically and produces photons through pion decays \citep{2011JCAP...03..051C, 2016ConPh..57..496G, 2024JCAP...03..035A}.
\\
\begin{figure}[!t]
    \centering
    \includegraphics[width=\linewidth]{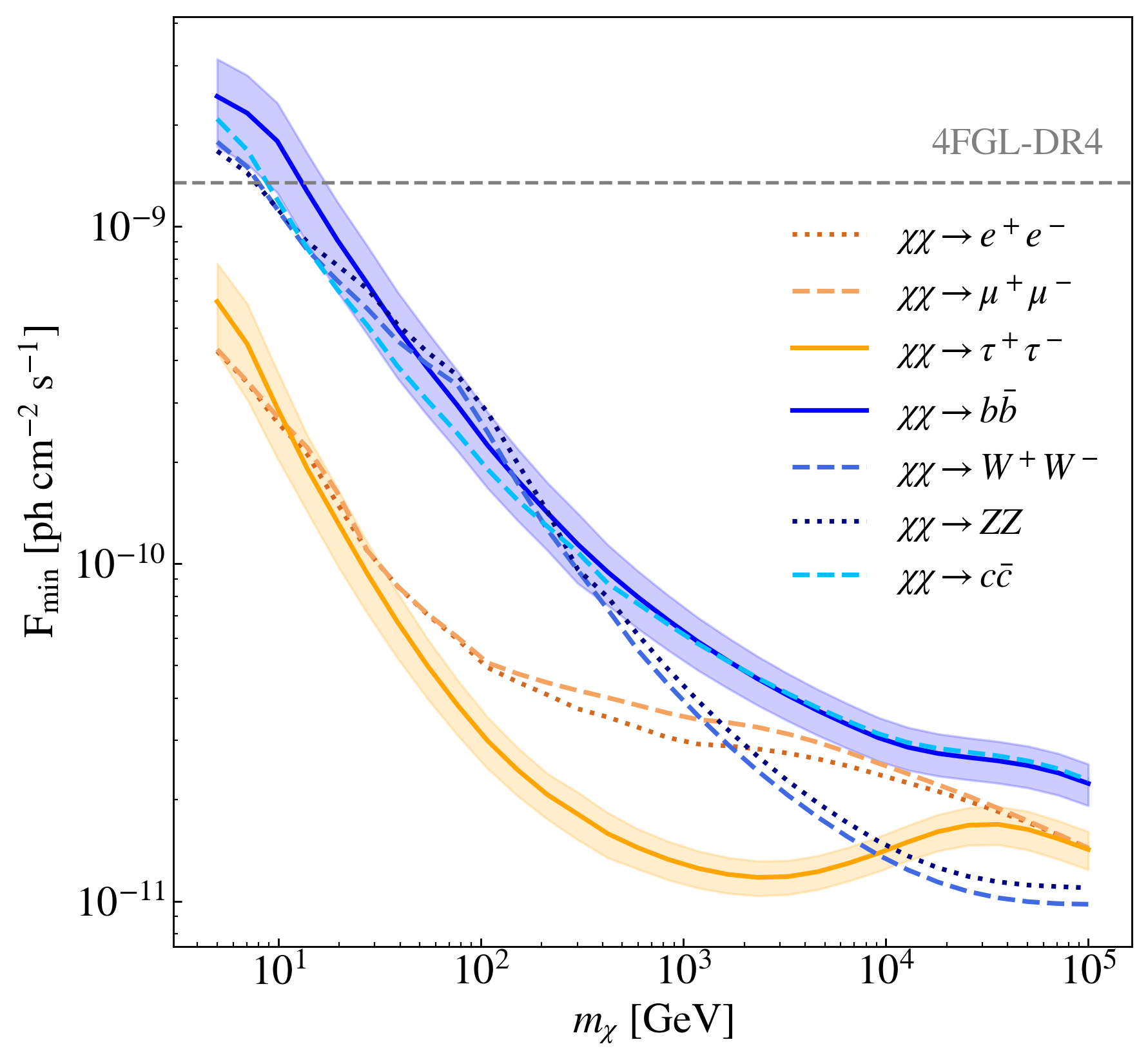}
    \caption{\textit{Fermi}-LAT sensitivity to dark satellites as a function of the DM mass, both for hadronic (blue) and leptonic  channels (orange). $1\sigma$ uncertainty bands resulting from averaging over the sky are included for the $b\bar{b}$ and $\tau^+\tau^-$ channels. The horizontal dashed line is the 4FGL-DR4 sensitivity threshold.
    }
    \label{fig-fmin}
\end{figure}
\begin{figure}
    \centering
    \includegraphics[width=\linewidth]{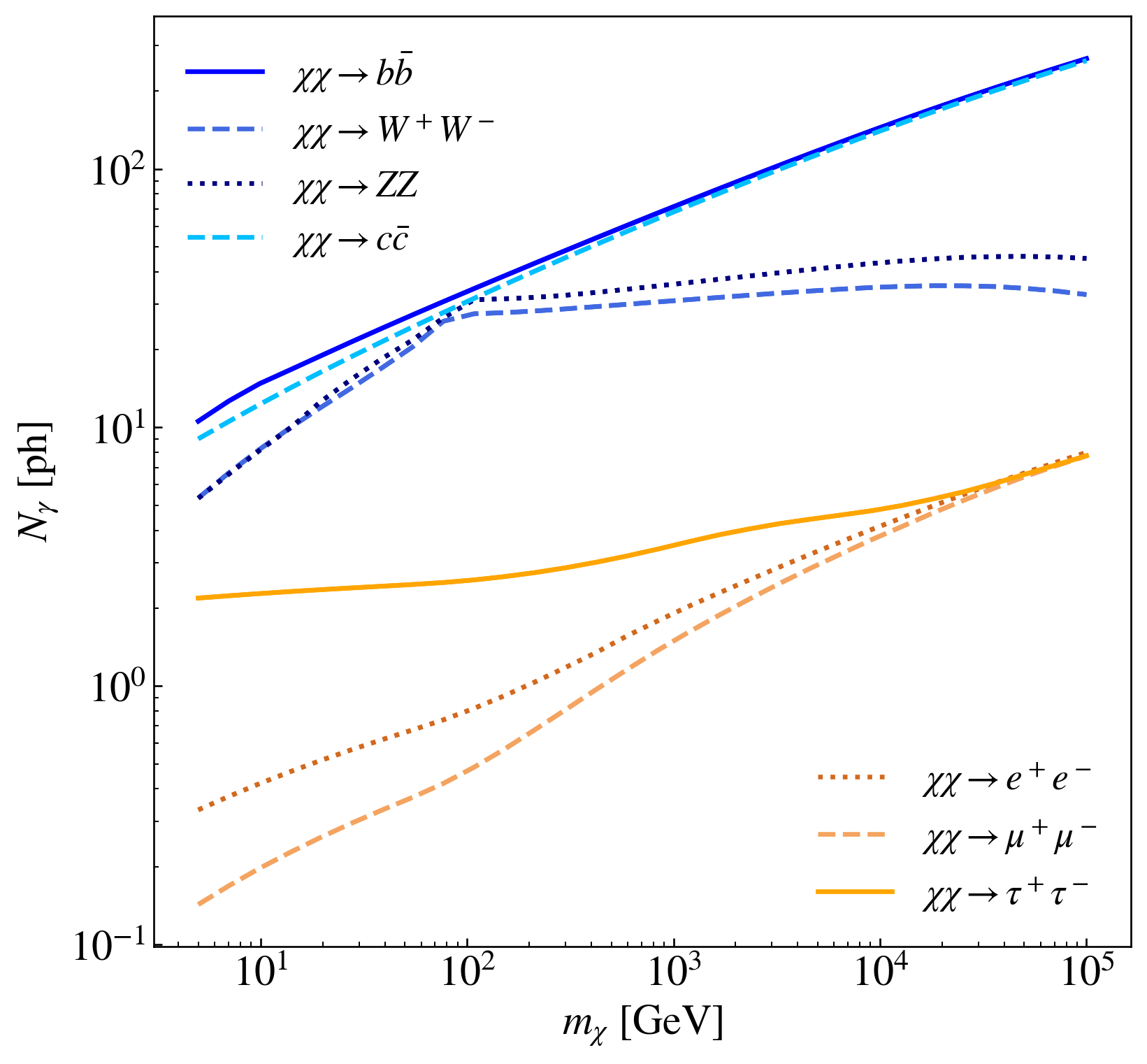}
    \caption{Number of gamma-ray photons produced by DM annihilation via different annihilation channels ---integrated DM annihilation spectra--- in the energy range covered by the 4FGL-DR4 catalog, i.e., $50$ MeV to 1 TeV.  
    We use DM mass values ranging from 5 GeV to 100 TeV. These results are computed with CosmiXs \citep{2024JCAP...03..035A}. Hadronic channels are represented in blue, while leptonic ones appear in orange.
    }
    \label{fig-ngamma}
\end{figure}

\noindent
\textit{C. Dark satellite J-factors:}
\vspace{1mm}

\noindent
The last ingredient in the derivation of the DM constraints is obtained from cosmological simulations, which offer valuable insights into the properties of the expected dark satellite Galactic population. We rely on the repopulated Via Lactea II (VL-II) simulation presented in \cite{2024MNRAS.530.2496A}, which extends the original DM-only VL-II subhalo population down to subhalo masses well below its resolution limit \citep{2008Natur.454..735D}. This enables the characterization of the full low-mass dark satellite population that is expected to be relevant in the context of \textit{Fermi}-LAT unIDs. We refer the reader to \cite{2024MNRAS.530.2496A} for the full details of the repopulation procedure\footnote{The repopulated data that we use can be found in the following repository: \href{https://projects.ift.uam-csic.es/damasco}{https://projects.ift.uam-csic.es/damasco}}. In that work, the position of the Sun is placed in 1000 random locations across the 1000 realizations of the repopulated halo, from which it is possible to draw statistically robust J-factor estimates and distributions. 

In our methodology, the specific J-factor value to be used for the derivation of the constraints following Eq. (\textcolor{blue}{\ref{eq-constraints}}) relies on the number of remaining unIDs in our list of possible dark satellite candidates, after the application of the different filters described in Section \textcolor{blue}{\ref{sec2}}. For instance, in Subsample I, where we are left with 220 unIDs, we would adopt the 220th largest J-factor value in each realization of the repopulated VL-II simulation. As $\langle \sigma v\rangle\propto J^{-1}$, we can see that the larger the J-factor, the stronger the constraints. We adopt a conservative approach in which, for each considered scenario with $n$ unIDs, we draw the J-factor distribution of the $n^{\rm th}$ brightest subhalo from the 1000 realizations, and adopt as our J-factor value the one above which 95$\%$ of such distribution is contained. Before this, we remove simulated subhalos with masses exceeding $10^7M_{\odot}$ to ensure that only dark satellites are considered. 

We obtain the following J-factor values for Subsamples I (220 unIDs), II (77 unIDs) and the SR case (1 unID): $9.9\times10^{17}$, $2.2\times10^{18}$ and $2.8\times10^{19}$ GeV$^2$cm$^{-5}$, respectively. Fig. \textcolor{blue}{\ref{fig-jfactor}} illustrates their corresponding distributions drawn from the 1000 Milky Way realizations. As expected, these distributions indicate that fainter subhalos exhibit a much smaller scatter in J-factor values across realizations, while the J-factor of the brightest subhalo varies by about two orders of magnitude, depending on the specific repopulation.

Note that a limitation of this approach to set DM limits is that, in the case of no unID being compatible with DM (SR scenario), we still need to adopt at least the J-factor of the brightest subhalo --- thus one surviving unID --- in order to use Eq. (\textcolor{blue}{\ref{eq-constraints}}).

\begin{figure}[t]
    \centering
    \includegraphics[width=\linewidth]{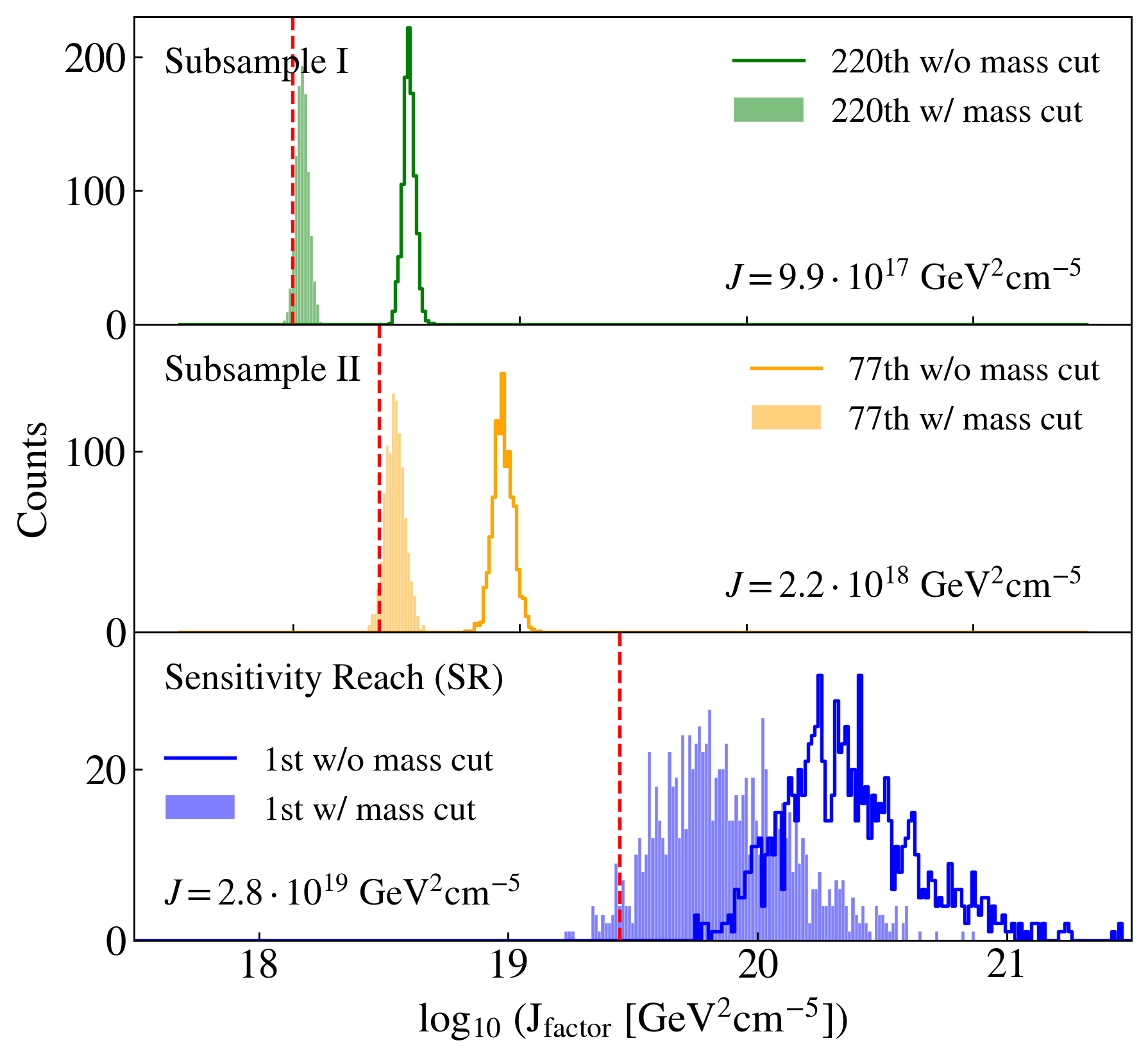}
    \caption{Distribution of the J-factor corresponding to the 220th (upper panel), 77th (middle panel) and 1st (lower panel) brightest subhalo in the 1000 realizations of the VL-II simulation performed in \cite{2024MNRAS.530.2496A}, before and after removing those DM subhalos with masses exceeding $10^7M_{\odot}$ (empty and filled histograms, respectively). For each scenario/panel, the quoted J-factor, also marked by the red dashed vertical line, is the one above which 95$\%$ of the distribution is contained.
    }
    \label{fig-jfactor}
\end{figure}

\subsection{DM upper limits}\label{Sec-5.2}

\noindent
Following the procedure outlined above, we apply Eq. (\ref{eq-constraints}) to set DM constraints for the three scenarios under consideration: Subsample I (220 sources), Subsample II (77 sources), and SR (no sources). Fig. \ref{fig-constraints} shows the 95$\%$ upper limits on the averaged annihilation cross section vs. DM mass, for both the $b\bar{b}$ and $\tau^+\tau^-$ channels. In all cases, the shaded bands in the figure correspond to 68\% containment (1$\sigma$) values, and come from the uncertainties on the $F_{\rm min}$ predictions (Section~\ref{Sec-5.1}). 
In the most conservative case, Subsample I, our $95\%$ upper limits 
are able to rule out $\langle\sigma v \rangle\sim 2\times10^{-25}\text{ cm}^{3}\text{ s}^{-1}$ around DM masses $m_\chi\sim10$ GeV.
Restricting the candidate sample to the 77 unIDs of Subsample II improves these limits by a factor of $\sim 2$ with respect to Subsample I, allowing us to reach values of $\langle\sigma v \rangle\sim 9\times10^{-26}\text{ cm}^{3}\text{ s}^{-1}$ around DM masses $m_\chi\sim10$ GeV.
Both scenarios are not able to reach the thermal relic cross section $\langle \sigma v \rangle\sim 2\times 10^{-26}\text{ cm}^{3}\text{ s}^{-1}$ \citep{2012PhRvD..86b3506S} for DM masses $m_{\chi}>5$ GeV. 
The maximum improvement, as expected, is achieved when none of the 4FGL-DR4 sources are considered compatible with a dark satellite (SR scenario, see Section \textcolor{blue}{\ref{Sec-4.2}}).
In this SR scenario, the limits could potentially improve significantly, by a factor of $\sim 25$ with respect to Subsample I, ruling out thermal WIMPs below 50 GeV in the $b\bar b$ channel, and below 30 GeV in the $\tau^+\tau^-$ channel. We remind the reader, however, that these constraints should be regarded as predictions rather than actual constraints until each of the 77 unIDs in Subsample II has been robustly ruled out as a potential dark satellite. While this may seem overly optimistic, it is supported by ongoing efforts to associate \textit{Fermi}-LAT unIDs with known astrophysical objects, as discussed in Section \textcolor{blue}{\ref{Sec-3.2}}, or to identify new astrophysical classes among the unIDs, e.g.~\cite{2023MNRAS.520.1348G, 2025arXiv250314584A, 2025arXiv250403543C}.

When comparing with previous work that used a similar methodology to set constraints with the sample of \textit{Fermi}-LAT unIDs \citep{2019JCAP...07..020C}, we find that our results have improved the constraints by a factor of $\sim 2$, as expected from the enhancement of the \textit{Fermi}-LAT sensitivity (codified in the new $F_{\rm min}$ computed specifically for the 4FGL-DR4 setup, compared to the previous $F_{\rm min}$ for the 3FGL).\footnote{We checked that the choice of the DM annihilation spectra simulator (PPPC4 or CosmiXs) 
does not significantly impact the constraints.}
This can be seen in Fig. \ref{fig-constraints}, where the blue dot-dashed line represents the SR scenario derived in \cite{2019JCAP...07..020C} after their filtering of 3FGL unIDs. 
As for the comparison of our results with the current best DM upper limits in the field, derived from the study of dSphs with \textit{Fermi}-LAT \citep{2026ApJ..1004L..33C}, 
we find our SR to be just a factor $\sim2$ less constraining, thus having the potential to achieve almost comparable constraints. In this sense, our results are highly complementary to these other methodologies and targets, and independent of them.

\begin{figure}[!t]
    \centering
    \includegraphics[width=\linewidth]{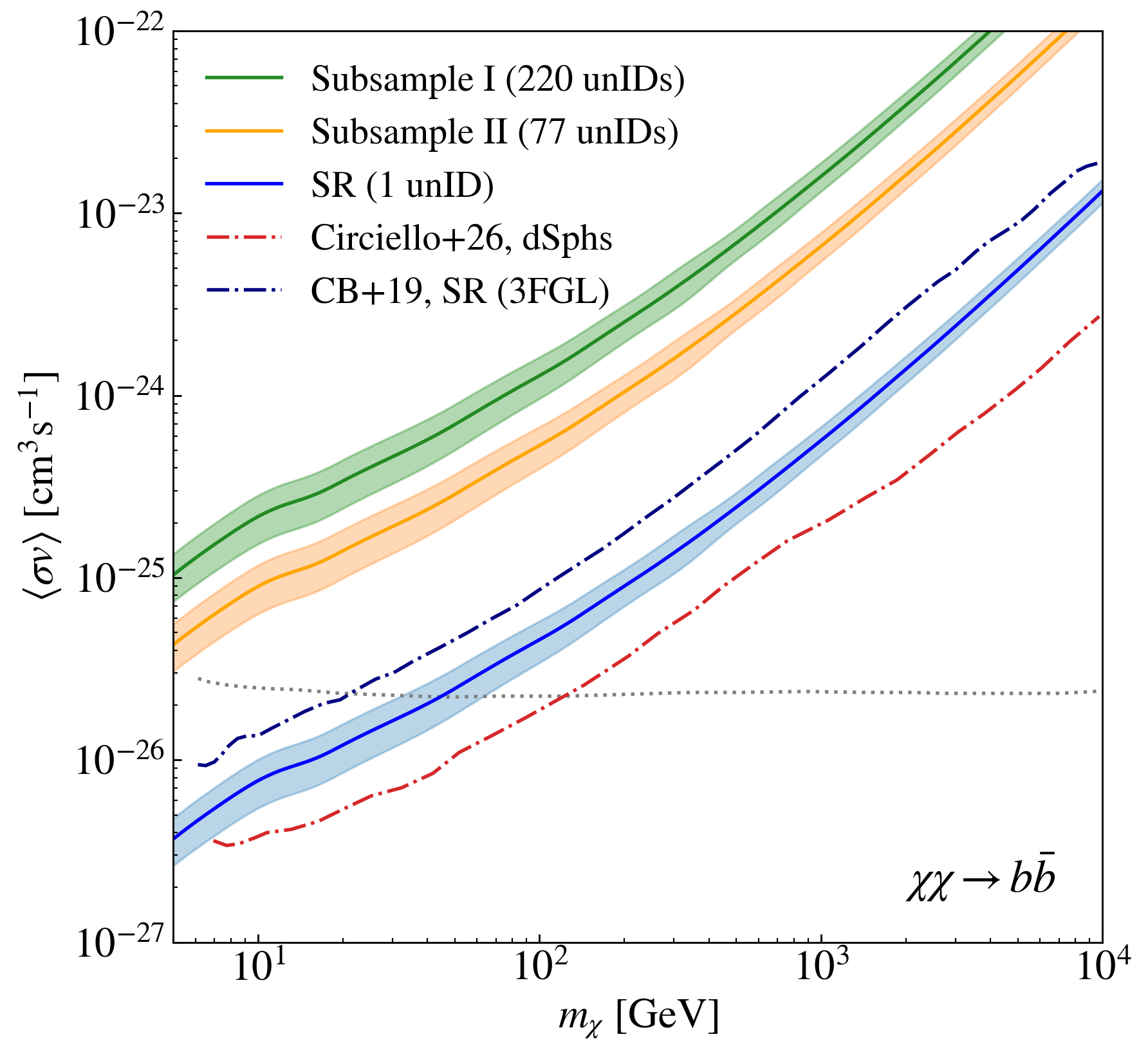}
    \includegraphics[width=\linewidth]{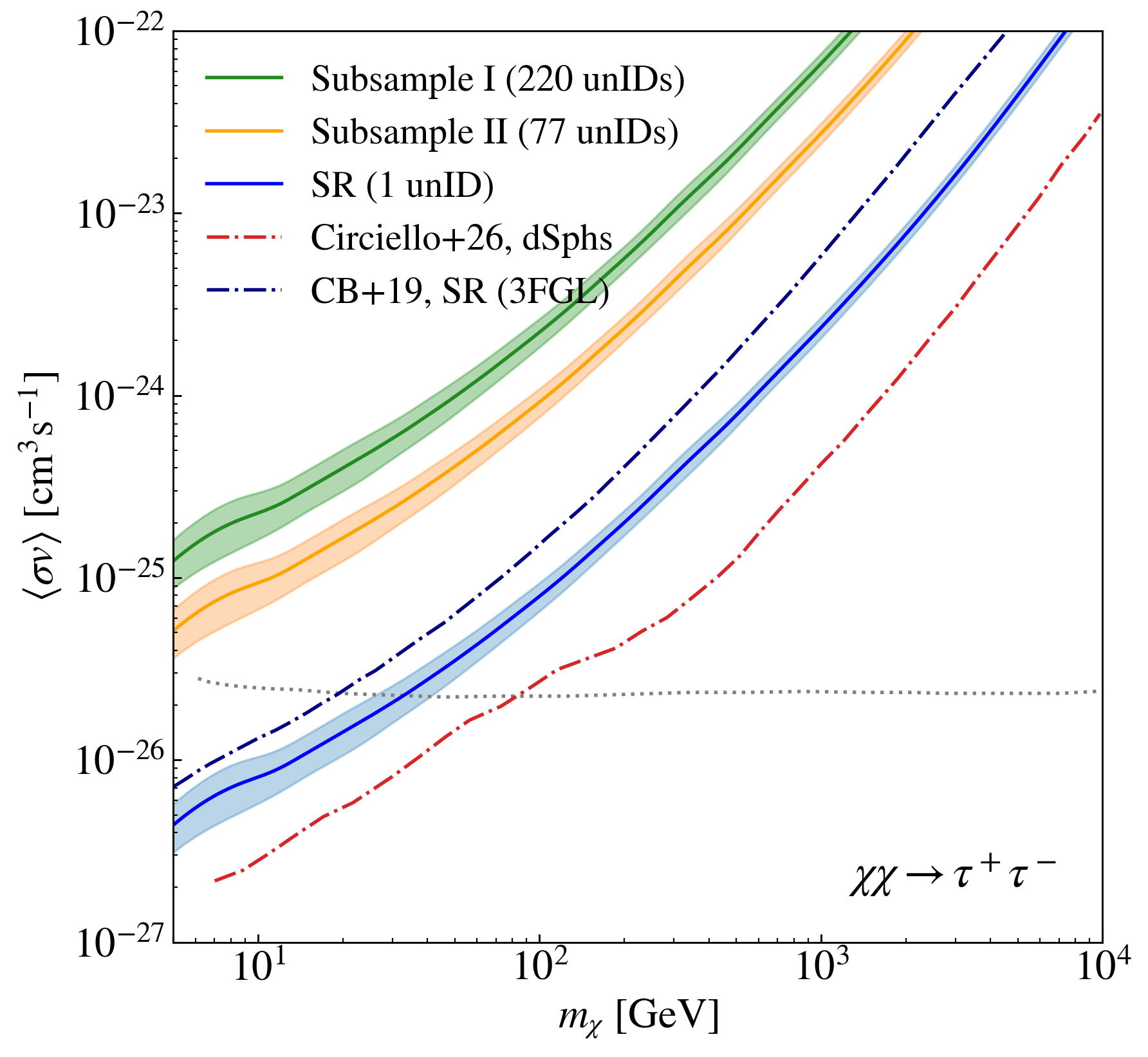}
    \caption{Constraints on the DM annihilation cross section for $b\bar{b}$ (top) and $\tau^+\tau^-$ (bottom) channels, for the three scenarios considered: Subsample I with 220 sources (green), Subsample II composed of 77 sources without any MW counterpart (orange), and the SR case in which no unID is compatible with DM (blue). The shaded regions represent the $1\sigma$ uncertainty band coming from the calculation of the $F_{\rm min}$; see Fig.~\ref{fig-fmin}. The gray dotted line represents the thermal annihilation cross section~\citep{2012PhRvD..86b3506S}.
    We compare with the DM limits derived from the \textit{Fermi}-LAT analysis of dSphs~\citep{2026ApJ..1004L..33C} (red dot-dashed line). The constraints obtained in previous work adopting the same methodology on the 3FGL catalog are also shown as a blue dot-dashed line for their SR scenario (\citealp{2019JCAP...07..020C}; CB+19). Compared with our SR case, our exclusion line is a factor $\sim2$ more constraining due to the improvement in $F_{\rm min}$; see text for details.
    }
    \label{fig-constraints}
\end{figure}

\section{Conclusions}\label{Sec-6}

\noindent
In this work, we have revisited the longstanding challenge of detecting DM through gamma rays. In particular, we addressed the question of whether the observed gamma-ray emission from the population of \textit{Fermi}-LAT unidentified sources (unIDs) could be explained as the signature of WIMP annihilation in nearby dark satellites, i.e., DM subhalos so light that they are devoid of baryonic content. Starting from the 14-year 4FGL-DR4 \textit{Fermi}-LAT catalog, we scrutinized its 2428 unIDs in search of those that could be compatible with being dark satellites. To do so, we first applied a series of selection filters that were designed considering the expected properties for both standard astrophysical emitters and predicted $\Lambda$CDM dark satellites. Our initial criteria included Galactic latitude, temporal variability and detection significance cuts (Section~\ref{Sec-3.1}), that were later complemented by a simple spectral analysis based on the so-called $\beta$-plot (Fig.~\ref{plot-subsamples}). The latter allows for a first-order separation between potential DM-induced gamma-ray emission --- in our case informed by the CosmiXs library for annihilation spectra --- versus other, well-known astrophysical object classes (pulsars, AGNs). This exercise led to a first subsample of 220 unIDs as potential dark satellites (`Subsample I'; see Section~\ref{sec-3.2}). Then, a careful multi-wavelength examination of these remaining unIDs was performed, which allowed us to further decrease the number of potential dark satellites down to 77 unIDs, i.e., those in Subsample I exhibiting no potential emissions at other wavelengths (`Subsample II'). 

For each of the 77 unIDs composing our Subsample II, a dedicated \textit{Fermi}-LAT spectral analysis was performed, revealing no statistically significant evidence in favor of DM annihilation (Section~\ref{Sec-4}). Indeed, our comparison of DM versus astrophysical spectral models, based on the Akaike Information Criterion, yielded only two sources reaching the nominal strong-evidence threshold (pre-trials). The resulting distribution of preferred DM masses and cross sections, though, showed no clustering or common patterns (see Fig.~\ref{fig:deltaaic}), thus disfavoring the hypothesis that a population of nearby dark satellites is responsible for the observed gamma-ray emission of such unIDs. 

In the absence of a convincing annihilation signal among the pool of unIDs, in Section~\ref{Sec-5} we derived upper limits on the thermally averaged annihilation cross section, by comparing the number of surviving dark satellite candidates in the catalog with predictions for the population of $\Lambda$CDM dark satellites in a Milky Way-size halo as obtained via numerical cosmological simulations, namely a `repopulated' version of the VL-II simulation \citep{2024MNRAS.530.2496A}. 
In addition to using a newer catalog and a different number of remaining unIDs, we note that, compared to previous work \citep{2019JCAP...07..020C,2019JCAP...11..045C}, our computation of DM constraints incorporates updated \textit{Fermi}-LAT sensitivity maps for each DM annihilation channel, state-of-the-art CosmiXs photon yields, and up-to-date VL-II J-factor estimates (see Section~\ref{Sec-5.1}). 
Our upper limits for the $b\bar{b}$ and $\tau^+\tau^-$ channels, obtained by adopting the number of unIDs in Subsamples I and II and summarized in Fig.~\ref{fig-constraints}, are a factor of $\sim$2--3 above the thermal relic cross section value at the lowest considered WIMP masses, and weaken quickly toward larger mass values as expected. 

Motivated by the absence of an unequivocal dark satellite among the whole pool of 4FGL-DR4 unIDs, we then considered an additional scenario in which none of the unIDs can be attributed to DM.
In this sensitivity-reach (SR) scenario, our predicted bounds become competitive with the latest constraints from dwarf spheroidal galaxies~\citep{2026ApJ..1004L..33C}, and exclude thermal WIMPs up to $\sim 50$ GeV ($b\bar{b}$) and $\sim 30$ GeV ($\tau^+\tau^-$). This represents an improvement of a factor of $\sim 2$ over previous SR predictions~\textcolor{blue}{\citep{2019JCAP...07..020C}}, primarily due to the increased exposure and energy reach of the 4FGL-DR4 catalog. We emphasize, however, that these results should be regarded as forecasted sensitivities rather than actual constraints, at least until each of the 77 unIDs in Subsample II can be robustly ruled out as a potential dark satellite. While this assumption may be viewed as optimistic, it is motivated by ongoing efforts to associate \textit{Fermi}-LAT unIDs with known astrophysical counterparts (Section \textcolor{blue}{\ref{Sec-3.2}}), as well as by recent studies uncovering new astrophysical source classes within the unID population, e.g.~\cite{2023MNRAS.520.1348G, 2025arXiv250314584A, 2025arXiv250403543C}.

Overall, our results reinforce the growing tension between the canonical WIMP paradigm and the continued absence of an accompanying gamma-ray signal. Yet, they also highlight the unique value of unID-based searches: they provide constraints that are complementary to those derived from classical DM targets such as dwarf spheroidals, and they probe a distinct, cosmologically motivated class of objects -- dark satellites -- that remain otherwise observationally inaccessible. 

Future perspectives include the scrutiny of new \textit{Fermi}-LAT catalogs and the incorporation of J-factors derived from hydrodynamical cosmological simulations. Ongoing efforts to implement baryonic physics processes in DM-only simulations predict up to an order-of-magnitude decrease in the number of subhalos in the innermost $\sim 25$ kpc of the Galaxy, where  dark satellites are most relevant for the purposes of this work (e.g., \citealp{2017MNRAS.471.1709G}).  
Recent work has begun addressing the effects of baryons and tidal stripping on the dark satellite population, finding it to be extremely resilient (e.g., \citealp{2023MNRAS.518...93A,2026PDU....5202262P, 2026MNRAS.545f2058A}). 
Yet, the debate is ongoing regarding the survival of dark satellites as well as their precise radial distribution in the innermost regions of Milky Way-size halos. This lack of consensus introduces a significant systematic uncertainty in the J-factor distributions. While our current estimates rely on DM-only simulations and do not explicitly model these effects, we are currently addressing this intrinsic systematic uncertainty, which will be presented in a dedicated follow-up study. 

On the other hand, current and future efforts towards a robust association of unIDs through MW observations or robust machine-learning techniques may clarify the origin of some of the dark satellite candidates presented here. 
Along these lines, future MeV instruments, such as the COSI mission \citep{2024icrc.confE.745T} or proposed concepts like AMEGO-X \citep{2022JATIS...8d4003C} and newASTROGAM \citep{2025arXiv250708133B}, could help identify MeV-peaked AGN counterparts among the current unID population, further reducing the pool of dark satellite candidates.
For the most compelling candidates that survive this scrutiny, complementary follow-up strategies exist beyond the gamma-ray band.
The future CTAO will enable detailed spectral follow-up observations of the best dark satellite candidates at higher energies. Meanwhile, independent gravitational probes such as pulsar timing \citep{2019PhRvD.100b3003D, 2026PhRvL.136d1201C} or microlensing could test the dark-satellite hypothesis, 
further enhancing the discovery potential of this approach.
\\

\noindent
\vspace{1cm}
\noindent\begin{minipage}{\linewidth}
\footnotesize
\setlength{\baselineskip}{0.92\baselineskip} 
\textit{Acknowledgements:} We would like to thank Cristina Fernández-Suárez and Pedro de la Torre-Luque for useful discussions.
FV and JMC acknowledge support from the European Union through the grant ``UNDARK'' 
of the Widening participation and spreading excellence programme (project number 101159929). 
FV and JMC also acknowledge the MICINN through the grant ``DarkMaps'' PID2022-142142NB-I00. The work of MASC was supported by the grants PID2024-155874NB-C21 and CEX2020-001007-S, both funded by MCIN/AEI/10.13039/501100011033 and by ``ERDF A way of making Europe''. The authors also acknowledge the MultiDark Network, ref. RED2022-134411-T.
The \textit{Fermi}-LAT Collaboration acknowledges generous ongoing support from a number of agencies
and institutes that have supported both the development and the operation of the LAT as well as
scientific data analysis. These include the National Aeronautics and Space Administration and the
Department of Energy in the United States, the Commissariat à l’Energie Atomique and the Centre
National de la Recherche Scientifique / Institut National de Physique Nucléaire et de Physique des
Particules in France, the Agenzia Spaziale Italiana and the Istituto Nazionale di Fisica Nucleare in
Italy, the Ministry of Education, Culture, Sports, Science and Technology (MEXT), High Energy Accelerator Research Organization (KEK) and Japan Aerospace Exploration Agency (JAXA) in Japan,
and the K. A. Wallenberg Foundation, the Swedish Research Council and the Swedish National Space
Board in Sweden. Additional support for science analysis during the operations phase is gratefully
acknowledged from the Istituto Nazionale di Astrofisica in Italy and the Centre National d’Etudes
Spatiales in France. This work was performed in part under DOE Contract DE-AC02-76SF00515.
The authors would like to thank Nestor Mirabal and Giacomo Principe for their helpful comments and suggestions during the internal review process.

\end{minipage}
\vspace{0.6em}

\bibliographystyle{elsarticle/elsarticle-harv} 

\bibliography{ref}


\newpage
\appendix

\onecolumn
\section{Rejection criteria for the Subsample I unIDs} \label{app-B}

\noindent
In Table \textcolor{blue}{\ref{table-rejection}} we summarize the rejection criteria for the 143 unIDs from Subsample I that exhibit MW emission. As detailed in Section \textcolor{blue}{\ref{Sec-3.2}}, dark satellites are expected to emit only gamma rays, thus, any additional emission would suggest that these subhalos have retained baryons and the DM searches would be more difficult on them. Therefore, to select those that are more suitable to be completely dark satellites without any astrophysical counterpart we have rejected those with another emission searching on SIMBAD, and recent literature efforts aimed at clarifying the nature of these objects. Note that for X-rays we have used Machine-Learning (ML) classifications of the X-ray counterpart candidates of \textit{Fermi}-LAT unIDs from CXO, eROSITA or Swift-XRT. In contrast, ML algorithms trained on \textit{Fermi}-LAT sources to classify the unIDs themselves from their gamma-ray properties are currently a subject of significant debate within the community, particularly regarding the potential for generating spurious results. Therefore, for the purpose of this work, and in order to be conservative, we have not included this second type of ML classification. Accurate implementation of these ML methods could potentially reduce the number of shortlisted sources, thereby refining the constraints we can place on the DM parameter space.

\section{Preferred channel of the best DM-fit}
In Table \textcolor{blue}{\ref{table-dmfit-aicc}} we summarize the preferred DM annihilation channel for the Subsample II sources, where a suitable DM fit is identified. The overall $\Delta$AIC$_c$ is shown in Fig. \textcolor{blue}{\ref{fig:deltaaic}}. Additionally, we present the TS of detection, the $\Delta$AIC$_c$ values, and the best-fit values for $\langle \sigma v\rangle $ and $m_{\chi}$ as well as the 1$\sigma$ errors.

\section{Promising candidates for follow-up}
\label{App-E}
In Section \textcolor{blue}{\ref{Sec-4}} we mentioned that two of the Subsample II unIDs reached the nominal strong-evidence threshold (pre-trials) in terms of the $\Delta$AIC$_c$ comparison with typical data-driven astrophysical models. These sources are: 4FGL J2342.4-4739 and 4FGL J1225.9+2951. Fig. \textcolor{blue}{\ref{fig:sed-4unids}} shows the corresponding SED fits. These candidates, among others, will be further examined in a follow-up work.

\setcounter{figure}{0}
\renewcommand{\thefigure}{C.\arabic{figure}}
\begin{figure}[!b]
    \centering
    \includegraphics[width=0.85\linewidth]{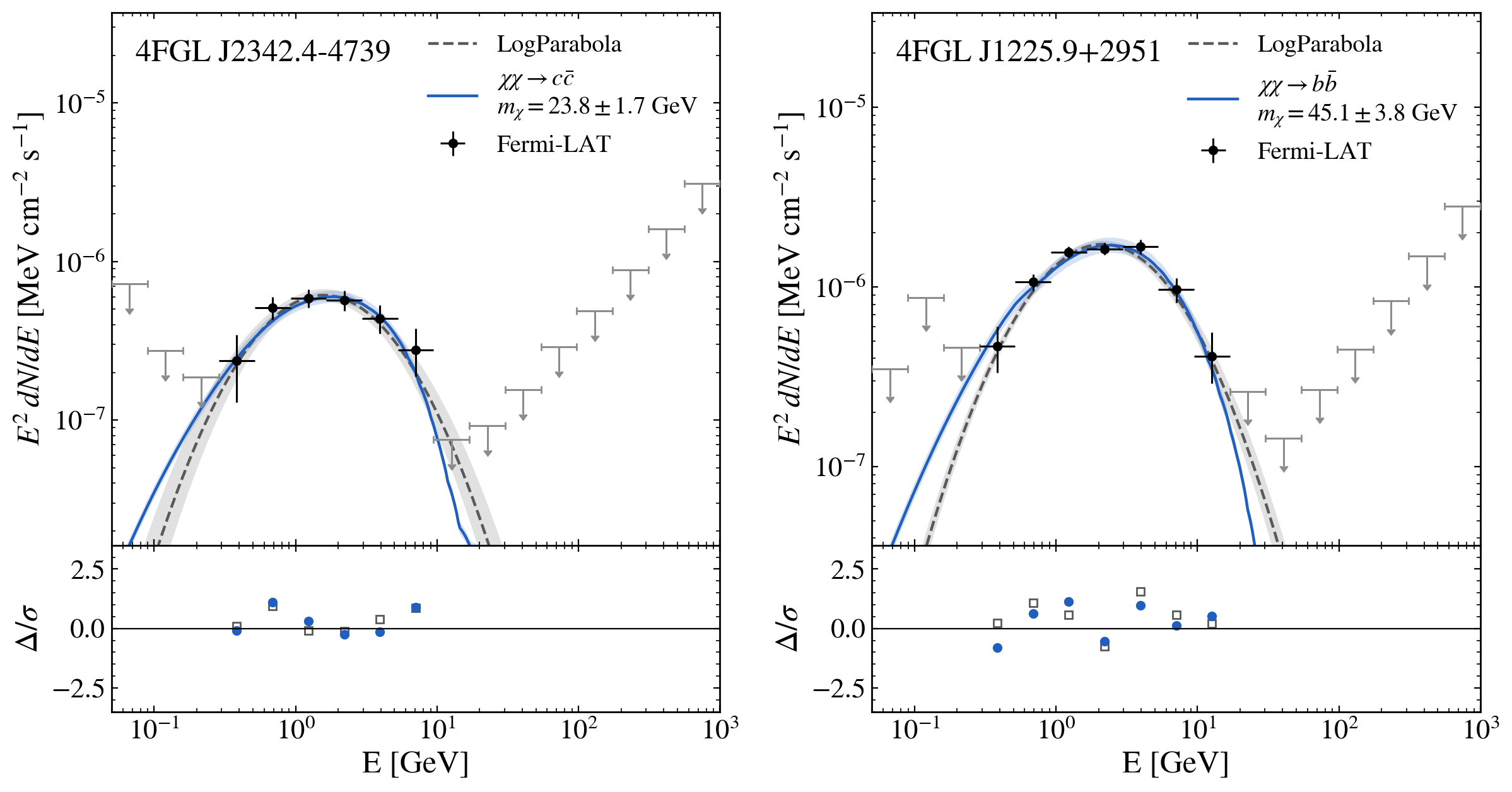}
    \caption{SED for the two unIDs with a $\Delta$AIC$_c$ below the nominal \textit{strong-evidence} threshold (pre-trials) for a DM fit over the astrophysical fits.}
    \label{fig:sed-4unids}
\end{figure}

\onecolumn

\setcounter{table}{0}           
\renewcommand{\thetable}{A.\arabic{table}}
\begin{small}
\begin{longtable}{p{0.22\textwidth} p{0.40\textwidth} p{0.30\textwidth}}
\caption{Rejection criteria for the 143 rejected unIDs from Subsample I. The listed rejection cause includes every emission or MW classification reported in the consulted literature. See Section~\ref{Sec-3.2} for details.}
\label{table-rejection}\\

\toprule
Source Name & Rejection Cause & Reference(s)\\
\midrule
\endfirsthead

\caption[]{\textit{(Continued)}}\\
\toprule
Source Name & Rejection Cause & Reference(s)\\
\midrule
\endhead

\midrule \multicolumn{3}{r}{\emph{Continues on next page}}\\ \midrule
\endfoot

\bottomrule
\endlastfoot

    4FGL J0003.3+2511 & Radio & \textcolor{blue}{\cite{2021ApJ...914...42B}} \\
    4FGL J0003.6+3059 & Radio & \textcolor{blue}{\cite{2021ApJ...914...42B}}\\
    4FGL J0026.1-0732 & X-ray, Radio, Blazar candidate & \textcolor{blue}{\cite{2021ApJ...923...75K}}, \textcolor{blue}{\cite{2021ApJ...914...42B}}, \textcolor{blue}{\href{http://simbad.cds.unistra.fr/simbad/sim-id?Ident=\%405224790\&Name=4FGL\%20J0026.1-0732\&submit=submit}{SIMBAD}}\\
    4FGL J0031.0-2327 & CLASS2 (agn), X-ray, Radio, BL Lac candidate & \textcolor{blue}{\cite{2023arXiv230712546B}}, \textcolor{blue}{\cite{2021ApJ...923...75K}}, \textcolor{blue}{\cite{2021ApJ...914...42B}}, \textcolor{blue}{\href{http://simbad.cds.unistra.fr/simbad/sim-id?Ident=4FGL+J0031.0-2327\&NbIdent=1&Radius=5&Radius.unit=arcmin&submit=submit+id}{SIMBAD}} \\
    4FGL J0031.5-5648 & X-ray & \textcolor{blue}{\cite{2021ApJ...923...75K}} \\
    4FGL J0040.2-2725 & CLASS2 (agn), Radio, Galaxy Cluster candidate & \textcolor{blue}{\cite{2021ApJ...914...42B}}, \textcolor{blue}{\cite{2023arXiv230712546B}} \\
    4FGL J0102.3+1000 & Radio & \textcolor{blue}{\cite{2021ApJ...914...42B}} \\
    4FGL J0106.1+1131 & Radio & \textcolor{blue}{\cite{2021ApJ...914...42B}} \\
    4FGL J0115.4-2917 & CLASS2 (agn), Radio, IR, AGN candidate & \textcolor{blue}{\cite{2023arXiv230712546B}}, \textcolor{blue}{\cite{2021ApJ...914...42B}}, \textcolor{blue}{\href{https://vizier.cds.unistra.fr/viz-bin/VizieR-S?WISEA\%20J011546.92-291703.6}{SIMBAD}}\\
    4FGL J0122.4+1034 & Radio & \textcolor{blue}{\cite{2021ApJ...914...42B}} \\
    4FGL J0125.9-6303 & X-ray & \textcolor{blue}{\cite{2021ApJ...923...75K}}, \textcolor{blue}{\href{https://simbad.cds.unistra.fr/simbad/sim-id?Ident=4FGL+J0125.9-6303&NbIdent=1&Radius=2&Radius.unit=arcmin&submit=submit+id}{SIMBAD}} \\
    4FGL J0126.3-6746 & X-ray & \textcolor{blue}{\cite{2021ApJ...923...75K}}\\
    4FGL J0137.3-3239 & X-ray, Radio, Blazar candidate & \textcolor{blue}{\cite{2021ApJ...923...75K}}, \textcolor{blue}{\cite{2023ApJ...943...51B}}, \textcolor{blue}{\cite{2021ApJ...914...42B}} \\
    4FGL J0139.5-2228 & Radio & \textcolor{blue}{\cite{2021ApJ...914...42B}}\\
    4FGL J0159.8-2234 & Radio & \textcolor{blue}{\cite{2021ApJ...914...42B}}\\
    4FGL J0251.1-1830 & X-ray, Radio & \textcolor{blue}{\cite{2024MNRAS.530.4626U}}, \textcolor{blue}{\cite{2021ApJ...914...42B}}, \textcolor{blue}{\href{https://simbad.cds.unistra.fr/simbad/sim-id?Ident=4FGL+J0251.1-1830&NbIdent=1&Radius=2&Radius.unit=arcmin&submit=submit+id}{SIMBAD}}\\
    4FGL J0301.6-5617 & X-ray & \textcolor{blue}{\cite{2021ApJ...923...75K}}\\
    4FGL J0312.7+2012 & Radio &  \textcolor{blue}{\cite{2021ApJ...914...42B}}, \textcolor{blue}{\href{https://simbad.cds.unistra.fr/simbad/sim-id?Ident=4FGL+J0312.7\%2B2012&NbIdent=1&Radius=2&Radius.unit=arcmin&submit=submit+id}{SIMBAD}} \\
    4FGL J0347.5+0722 & Radio & \textcolor{blue}{\cite{2021ApJ...914...42B}}\\
    4FGL J0409.4-3822 & Radio & \textcolor{blue}{\cite{2021ApJ...914...42B}}\\
    4FGL J0433.0+1327 & Radio & \textcolor{blue}{\cite{2021ApJ...914...42B}} \\
    4FGL J0438.0-7329 & IR, BL Lac candidate & \textcolor{blue}{\href{https://simbad.cds.unistra.fr/simbad/sim-id?Ident=4FGL+J0438.0-7329&NbIdent=1&Radius=2&Radius.unit=arcmin&submit=submit+id}{SIMBAD}} \\
    4FGL J0459.7-5413 & Blazar candidate &  \textcolor{blue}{\href{https://simbad.cds.unistra.fr/simbad/sim-id?Ident=4FGL+J0459.7-5413&NbIdent=1&Radius=2&Radius.unit=arcmin&submit=submit+id}{SIMBAD}}\\
    4FGL J0536.1-1205 & CLASS2 (agn), Radio, IR, AGN candidate & \textcolor{blue}{\cite{2023arXiv230712546B}}, \textcolor{blue}{\cite{2021ApJ...914...42B}}, \textcolor{blue}{\href{https://vizier.cds.unistra.fr/viz-bin/VizieR-S?WISEA\%20J053541.10-120244.2}{SIMBAD}}\\
    4FGL J0550.5-4356 & Blazar candidate &  \textcolor{blue}{\href{https://simbad.cds.unistra.fr/simbad/sim-id?Ident=4FGL+J0550.5-4356&NbIdent=1&Radius=2&Radius.unit=arcmin&submit=submit+id}{SIMBAD}}\\
    4FGL J0553.9-5048 & X-ray & \textcolor{blue}{\cite{2021ApJ...923...75K}}\\
    4FGL J0624.7-4903 & X-ray & \textcolor{blue}{\cite{2021ApJ...923...75K}}, \textcolor{blue}{\href{https://simbad.cds.unistra.fr/simbad/sim-id?Ident=4FGL+J0624.7-4903&NbIdent=1&Radius=2&Radius.unit=arcmin&submit=submit+id}{SIMBAD}} \\
    4FGL J0625.5+7029 & Radio & \textcolor{blue}{\cite{2021ApJ...914...42B}} \\
    4FGL J0650.2-5144 & Radio, Blazar candidate & \textcolor{blue}{\href{http://simbad.cds.unistra.fr/simbad/sim-id?Ident=4FGL+J0650.2-5144&NbIdent=1&Radius=5&Radius.unit=arcmin&submit=submit+id}{SIMBAD}} \\
    4FGL J0731.0+6607 & Radio & \textcolor{blue}{\cite{2021ApJ...914...42B}}\\
    4FGL J0755.6+1855 & Radio & \textcolor{blue}{\cite{2021ApJ...914...42B}}\\
    4FGL J0800.5+3715 & CLASS2 (agn), Radio, IR, AGN candidate & \textcolor{blue}{\cite{2023arXiv230712546B}}, \textcolor{blue}{\cite{2021ApJ...914...42B}}, \textcolor{blue}{\href{https://vizier.cds.unistra.fr/viz-bin/VizieR-S?WISEA\%20J080040.50\%2b371825.2}{SIMBAD}} \\
    4FGL J0807.2+3312 & Radio & \textcolor{blue}{\cite{2021ApJ...914...42B}}\\
    4FGL J0827.6-0059 & CLASS2 (unk), Radio & \textcolor{blue}{\cite{2023arXiv230712546B}}, \textcolor{blue}{\cite{2021ApJ...914...42B}} \\
    4FGL J0906.1-1011 & X-ray, Radio & \textcolor{blue}{\href{http://simbad.cds.unistra.fr/simbad/sim-id?Ident=4FGL+J0906.1-1011&NbIdent=1&Radius=5&Radius.unit=arcmin&submit=submit+id}{SIMBAD}}, \textcolor{blue}{\cite{2021ApJ...914...42B}}, \textcolor{blue}{\cite{2023ApJ...943...51B}}\\
    4FGL J0910.9+6055 & Radio & \textcolor{blue}{\cite{2021ApJ...914...42B}}\\
    4FGL J0942.5+7620 & Radio & \textcolor{blue}{\cite{2021ApJ...914...42B}}\\
    4FGL J0949.8+7551 & Radio & \textcolor{blue}{\cite{2021ApJ...914...42B}}\\
    4FGL J0953.6-1509 & X-ray & \textcolor{blue}{\cite{2024A&A...684A.208M}} \\
    4FGL J0958.6-0522 & Radio & \textcolor{blue}{\cite{2021ApJ...914...42B}}\\
    4FGL J1001.3-2040 & Radio & \textcolor{blue}{\cite{2021ApJ...914...42B}}\\
    4FGL J1009.5-0034 & Radio & \textcolor{blue}{\cite{2021ApJ...914...42B}}\\
    4FGL J1014.6+6126 & X-ray & \textcolor{blue}{\cite{2021ApJ...923...75K}}\\
    4FGL J1018.2+0145  & X-ray & \textcolor{blue}{\cite{2024ApJ...971..180Y}}\\
    4FGL J1019.3+5625 & Radio & \textcolor{blue}{\cite{2021ApJ...914...42B}}, \textcolor{blue}{\href{http://simbad.cds.unistra.fr/simbad/sim-id?Ident=4FGL+J1019.3\%2B5625+&NbIdent=1&Radius=5&Radius.unit=arcmin&submit=submit+id}{SIMBAD}}\\
    4FGL J1028.4-1539  & Radio & \textcolor{blue}{\cite{2021ApJ...914...42B}}\\
    4FGL J1029.9+7335  & Radio & \textcolor{blue}{\cite{2021ApJ...914...42B}}\\
    4FGL J1047.2+6740  & Radio, X-ray, Blazar candidate & \textcolor{blue}{\cite{2021ApJ...914...42B}}, \textcolor{blue}{\cite{2023ApJ...943...51B}}, \textcolor{blue}{\href{http://simbad.cds.unistra.fr/simbad/sim-id?Ident=4FGL+J1047.2\%2B6740&NbIdent=1&Radius=5&Radius.unit=arcmin&submit=submit+id}{SIMBAD}}\\
    4FGL J1059.4+0226 & Radio & \textcolor{blue}{\cite{2021ApJ...914...42B}}\\
    4FGL J1100.0-2044  & Radio & \textcolor{blue}{\cite{2021ApJ...914...42B}}\\
    4FGL J1102.3+6715 & CLASS2 (agn), Radio, IR & \textcolor{blue}{\cite{2023arXiv230712546B}}, \textcolor{blue}{\cite{2021ApJ...914...42B}}, \textcolor{blue}{\href{https://vizier.cds.unistra.fr/viz-bin/VizieR-S?WISEA\%20J110233.85\%2b671108.9}{SIMBAD}}\\
    4FGL J1106.7-1742 & Radio & \textcolor{blue}{\cite{2021ApJ...914...42B}}\\
    4FGL J1131.6+4657 & X-ray & \textcolor{blue}{\cite{2024MNRAS.530.4626U}}\\
    4FGL J1140.4+0022 & CLASS2 (agn), IR & \textcolor{blue}{\cite{2023arXiv230712546B}}, \textcolor{blue}{\href{https://vizier.cds.unistra.fr/viz-bin/VizieR-S?WISEA\%20J114039.87\%2b001958.0}{SIMBAD}}\\
    4FGL J1146.0-0638 & X-ray, Radio, BL Lac candidate & \textcolor{blue}{\cite{2024MNRAS.530.4626U}}, \textcolor{blue}{\cite{2021ApJ...914...42B}}, \textcolor{blue}{\href{http://simbad.cds.unistra.fr/simbad/sim-id?Ident=4FGL+J1146.0-0638+&NbIdent=1&Radius=5&Radius.unit=arcmin&submit=submit+id}{SIMBAD}}\\
    4FGL J1155.2-1111 & X-ray, Radio, BL Lac candidate & \textcolor{blue}{\cite{2021ApJ...923...75K}}, \textcolor{blue}{\cite{2021ApJ...914...42B}}, \textcolor{blue}{\href{http://simbad.cds.unistra.fr/simbad/sim-id?Ident=4FGL+J1155.2-1111+&NbIdent=1&Radius=5&Radius.unit=arcmin&submit=submit+id}{SIMBAD}}\\
    4FGL J1203.5-1748 & Radio & \textcolor{blue}{\cite{2021ApJ...914...42B}}, \textcolor{blue}{\cite{2023ApJ...943...51B}}\\
    4FGL J1217.2-2500  & Radio & \textcolor{blue}{\cite{2021ApJ...914...42B}}\\
    4FGL J1231.3+3353 & CLASS2 (agn), IR, AGN & \textcolor{blue}{\cite{2023arXiv230712546B}}, \textcolor{blue}{\href{https://vizier.cds.unistra.fr/viz-bin/VizieR-S?WISEA\%20J123119.47\%2b335244.7}{SIMBAD}}\\
    4FGL J1243.5+5311 & Radio & \textcolor{blue}{\cite{2021ApJ...914...42B}}\\
    4FGL J1308.7+0347 & CLASS2 (agn), X-ray, Radio & \textcolor{blue}{\cite{2023arXiv230712546B}}, \textcolor{blue}{\cite{2024ApJ...971..180Y}}, \textcolor{blue}{\cite{2021ApJ...914...42B}}\\
    4FGL J1331.3-3818 & Radio & \textcolor{blue}{\cite{2021ApJ...914...42B}}\\
    4FGL J1335.6+4759 & CLASS2 (agn), IR, AGN & \textcolor{blue}{\cite{2023arXiv230712546B}}, \textcolor{blue}{\href{https://vizier.cds.unistra.fr/viz-bin/VizieR-S?WISEA\%20J133534.14\%2b480228.0}{SIMBAD}}\\
    4FGL J1350.9-2757 & CLASS2 (unk), Radio & \textcolor{blue}{\cite{2023arXiv230712546B}}, \textcolor{blue}{\cite{2021ApJ...914...42B}}\\
    4FGL J1407.7-3017 & CLASS2 (unk), Radio, IR & \textcolor{blue}{\cite{2023arXiv230712546B}}, \textcolor{blue}{\cite{2021ApJ...914...42B}}, \textcolor{blue}{\cite{2023ApJ...943...51B}}, \textcolor{blue}{\href{https://vizier.cds.unistra.fr/viz-bin/VizieR-S?WISEA\%20J140826.40-302231.5}{SIMBAD}}\\
    4FGL J1408.7-3747 & Radio & \textcolor{blue}{\cite{2021ApJ...914...42B}}\\
    4FGL J1410.7+7405 & X-ray, Radio & \textcolor{blue}{\cite{2024MNRAS.530.4626U}}, \textcolor{blue}{\href{https://simbad.cds.unistra.fr/simbad/sim-id?Ident=4FGL+J1410.7\%2B7405+&NbIdent=1&Radius=2&Radius.unit=arcmin&submit=submit+id}{SIMBAD}}, \textcolor{blue}{\cite{2021ApJ...914...42B}}\\
   4FGL J1421.4-1655 & Radio & \textcolor{blue}{\cite{2021ApJ...914...42B}}\\
   4FGL J1429.8-0739 & X-ray, Radio & \textcolor{blue}{\cite{2021ApJ...923...75K}}, \textcolor{blue}{\cite{2021ApJ...914...42B}}, \textcolor{blue}{\href{http://simbad.cds.unistra.fr/simbad/sim-id?Ident=4FGL+J1429.8-0739&NbIdent=1&Radius=2&Radius.unit=arcmin&submit=submit+id}{SIMBAD}}\\
   4FGL J1441.4-1934 & X-ray, Radio  & \textcolor{blue}{\cite{2021ApJ...923...75K}}, \textcolor{blue}{\href{http://simbad.cds.unistra.fr/simbad/sim-id?Ident=4FGL+J1441.4-1934+&NbIdent=1&Radius=2&Radius.unit=arcmin&submit=submit+id}{SIMBAD}}\\
   4FGL J1442.0+0732 & CLASS2 (agn), Radio, IR, AGN candidate & \textcolor{blue}{\cite{2023arXiv230712546B}}, \textcolor{blue}{\cite{2021ApJ...914...42B}}, \textcolor{blue}{\href{https://vizier.cds.unistra.fr/viz-bin/VizieR-S?WISEA\%20J144149.00\%2b072911.6}{SIMBAD}}\\
   4FGL J1442.1-2800 & Radio & \textcolor{blue}{\cite{2021ApJ...914...42B}}\\
   4FGL J1442.1-3707 & Radio & \textcolor{blue}{\cite{2021ApJ...914...42B}}\\
   4FGL J1444.5+1922 & CLASS2 (agn), IR, AGN & \textcolor{blue}{\cite{2023arXiv230712546B}}, \textcolor{blue}{\href{https://vizier.cds.unistra.fr/viz-bin/VizieR-S?WISEA\%20J144442.59\%2b191750.9}{SIMBAD}}\\
   4FGL J1444.8+3645 & Radio & \textcolor{blue}{\cite{2021ApJ...914...42B}}\\
   4FGL J1450.8-1424 & Radio & \textcolor{blue}{\cite{2021ApJ...914...42B}}\\
   4FGL J1458.8-2120 & Radio & \textcolor{blue}{\cite{2021ApJ...914...42B}}\\
   4FGL J1500.2+2640 & Radio & \textcolor{blue}{\cite{2021ApJ...914...42B}}\\
   4FGL J1504.6+4343 & Radio & \textcolor{blue}{\cite{2021ApJ...914...42B}}\\
   4FGL J1510.9+0551 & Radio & \textcolor{blue}{\cite{2021ApJ...914...42B}}\\
   4FGL J1513.0-3118 & X-ray & \textcolor{blue}{\href{https://vizier.cds.unistra.fr/viz-bin/VizieR-4}{SIMBAD}}\\
   4FGL J1513.7-1519 & X-ray, Radio & \textcolor{blue}{\cite{2024A&A...684A.208M}}, \textcolor{blue}{\cite{2021ApJ...914...42B}}\\
   4FGL J1516.4-2640 & Radio & \textcolor{blue}{\cite{2021ApJ...914...42B}}\\
   4FGL J1519.7+6727 & X-ray, Radio & \textcolor{blue}{\href{https://vizier.cds.unistra.fr/viz-bin/VizieR-4}{SIMBAD}}, \textcolor{blue}{\cite{2021ApJ...914...42B}}\\
   4FGL J1527.8+1013 & Radio & \textcolor{blue}{\cite{2021ApJ...914...42B}}\\
   4FGL J1530.0-1522 & Radio & \textcolor{blue}{\cite{2021ApJ...914...42B}}\\
   4FGL J1532.9-0313 & Radio & \textcolor{blue}{\cite{2021ApJ...914...42B}}\\
   4FGL J1543.1+4209 & Radio & \textcolor{blue}{\cite{2021ApJ...914...42B}}\\
   4FGL J1543.6-0244 & Radio & \textcolor{blue}{\cite{2021ApJ...914...42B}}\\
   4FGL J1544.2-2554 & X-ray & \textcolor{blue}{\cite{2024A&A...684A.208M}}\\
   4FGL J1545.5+2839 & CLASS2 (agn), Radio, IR & \textcolor{blue}{\cite{2023arXiv230712546B}}, \textcolor{blue}{\cite{2021ApJ...914...42B}}, \textcolor{blue}{\href{https://vizier.cds.unistra.fr/viz-bin/VizieR-S?WISEA\%20J154520.76\%2b283508.6}{SIMBAD}}\\
   4FGL J1554.2+2008 & CLASS2 (agn), X-ray, Radio, AGN candidate & \textcolor{blue}{\cite{2023arXiv230712546B}}, \textcolor{blue}{\cite{2024MNRAS.530.4626U}}, \textcolor{blue}{\cite{2021ApJ...914...42B}}, \textcolor{blue}{\href{http://simbad.cds.unistra.fr/simbad/sim-id?Ident=4FGL+J1554.2\%2B2008&NbIdent=1&Radius=2&Radius.unit=arcmin&submit=submit+id}{SIMBAD}}\\
   4FGL J1555.2+7103 & Radio & \textcolor{blue}{\cite{2021ApJ...914...42B}}\\
   4FGL J1603.6-0451 & CLASS2 (sey), Radio & \textcolor{blue}{\cite{2023arXiv230712546B}}, \textcolor{blue}{\cite{2021ApJ...914...42B}}\\
   4FGL J1612.1+1407 & X-ray, Radio & \textcolor{blue}{\href{http://simbad.cds.unistra.fr/simbad/sim-id?Ident=4FGL+J1612.1\%2B1407+&NbIdent=1&Radius=2&Radius.unit=arcmin&submit=submit+id}{SIMBAD}}, \textcolor{blue}{\cite{2021ApJ...914...42B}}\\
   4FGL J1615.3-1512 & Radio & \textcolor{blue}{\cite{2021ApJ...914...42B}}\\
   4FGL J1626.5+6257 & CLASS2 (agn), IR, Radio, BL Lac candidate & \textcolor{blue}{\cite{2023arXiv230712546B}}, \textcolor{blue}{\href{http://simbad.cds.unistra.fr/simbad/sim-id?Ident=4FGL+J1626.5\%2B6257+&NbIdent=1&Radius=2&Radius.unit=arcmin&submit=submit+id}{SIMBAD}}\\
   4FGL J1631.8+4144 & X-ray, Radio & \textcolor{blue}{\cite{2024MNRAS.530.4626U}}, \textcolor{blue}{\href{http://simbad.cds.unistra.fr/simbad/sim-id?Ident=4FGL+J1631.8\%2B4144+&NbIdent=1&Radius=2&Radius.unit=arcmin&submit=submit+id}{SIMBAD}}, \textcolor{blue}{\cite{2021ApJ...914...42B}}\\
   4FGL J1632.5+3839 & Radio & \textcolor{blue}{\cite{2021ApJ...914...42B}}\\
   4FGL J1644.8+1850 & CLASS2 (agn), X-ray, Radio & \textcolor{blue}{\cite{2023arXiv230712546B}}, \textcolor{blue}{\href{http://simbad.cds.unistra.fr/simbad/sim-id?Ident=4FGL+J1644.8\%2B1850+&NbIdent=1&Radius=2&Radius.unit=arcmin&submit=submit+id}{SIMBAD}}, \textcolor{blue}{\cite{2021ApJ...914...42B}}\\
   4FGL J1648.7+4834 & X-ray, Radio & \textcolor{blue}{\cite{2024MNRAS.530.4626U}}, \textcolor{blue}{\cite{2023ApJ...943...51B}}, \textcolor{blue}{\cite{2021ApJ...914...42B}}, \textcolor{blue}{\href{http://simbad.cds.unistra.fr/simbad/sim-id?Ident=4FGL+J1648.7\%2B4834+&NbIdent=1&Radius=2&Radius.unit=arcmin&submit=submit+id}{SIMBAD}}\\
   4FGL J1653.0+3640 & Radio & \textcolor{blue}{\cite{2021ApJ...914...42B}}\\
   4FGL J1700.0-0122 & Radio & \textcolor{blue}{\cite{2021ApJ...914...42B}}\\
   4FGL J1710.0+4629 & Radio & \textcolor{blue}{\cite{2021ApJ...914...42B}}\\
    4FGL J1714.6+3228 & Radio & \textcolor{blue}{\cite{2021ApJ...914...42B}}\\
    4FGL J1716.1+2309 & Radio & \textcolor{blue}{\cite{2021ApJ...914...42B}}\\
    4FGL J1717.8+3925 & Radio & \textcolor{blue}{\cite{2021ApJ...914...42B}}\\
    4FGL J1720.6+0708 & Radio & \textcolor{blue}{\cite{2021ApJ...914...42B}}\\
    4FGL J1727.4+0326 & Radio & \textcolor{blue}{\cite{2021ApJ...914...42B}}\\
    4FGL J1731.6+0630 & Radio & \textcolor{blue}{\cite{2021ApJ...914...42B}}\\
    4FGL J1733.4+2235 & X-ray, Radio & \textcolor{blue}{\cite{2021ApJ...923...75K}}, \textcolor{blue}{\href{http://simbad.cds.unistra.fr/simbad/sim-id?Ident=4FGL+J1733.4\%2B2235+&NbIdent=1&Radius=2&Radius.unit=arcmin&submit=submit+id}{SIMBAD}}, \textcolor{blue}{\cite{2021ApJ...914...42B}}\\
    4FGL J1802.4+3634 & Radio & \textcolor{blue}{\cite{2021ApJ...914...42B}}\\
    4FGL J1813.5+2819 & Radio &\textcolor{blue}{\cite{2023ApJ...943...51B}}, \textcolor{blue}{\cite{2021ApJ...914...42B}} \\
   4FGL J1815.1+5917 & Radio & \textcolor{blue}{\cite{2021ApJ...914...42B}}\\
   4FGL J1918.6-7813c & CLASS2 (agn), Radio, Blazar candidate & \textcolor{blue}{\cite{2023arXiv230712546B}}, \textcolor{blue}{\href{http://simbad.cds.unistra.fr/simbad/sim-id?Ident=4FGL+J1918.6-7813c&NbIdent=1&Radius=2&Radius.unit=arcmin&submit=submit+id}{SIMBAD}}\\
   4FGL J1940.2-2511 & Radio & \textcolor{blue}{\cite{2023ApJ...943...51B}}, \textcolor{blue}{\cite{2021ApJ...914...42B}}\\
   4FGL J1959.3-6121 & Radio & \textcolor{blue}{\cite{2021ApJ...914...42B}}\\
   4FGL J2009.9-2116 & Radio & \textcolor{blue}{\cite{2021ApJ...914...42B}}\\
   4FGL J2013.9-8717 & Radio & \textcolor{blue}{\cite{2021ApJ...914...42B}}\\
   4FGL J2018.0+7903 & Radio & \textcolor{blue}{\cite{2021ApJ...914...42B}}\\
   4FGL J2023.3-1947 & Radio & \textcolor{blue}{\cite{2021ApJ...914...42B}}\\
   4FGL J2042.1-5320 & Radio & \textcolor{blue}{\cite{2021ApJ...914...42B}}\\
   4FGL J2046.1-1626 & Radio & \textcolor{blue}{\cite{2021ApJ...914...42B}}\\
   4FGL J2046.9-5409 & X-ray & \textcolor{blue}{\cite{2021ApJ...923...75K}}, \textcolor{blue}{\href{http://simbad.cds.unistra.fr/simbad/sim-id?Ident=4FGL+J2046.9-5409&NbIdent=1&Radius=2&Radius.unit=arcmin&submit=submit+id}{SIMBAD}}\\
   4FGL J2050.4-3015 & Radio & \textcolor{blue}{\cite{2021ApJ...914...42B}}\\
    4FGL J2054.1-4054 & CLASS2 (agn), Radio & \textcolor{blue}{\cite{2023arXiv230712546B}}, \textcolor{blue}{\cite{2021ApJ...914...42B}}\\
    4FGL J2056.4-5922 & Radio & \textcolor{blue}{\cite{2021ApJ...914...42B}}\\
    4FGL J2112.5-3043 & X-ray & \textcolor{blue}{\href{http://simbad.cds.unistra.fr/simbad/sim-id?Ident=4FGL+J2112.5-3043+&NbIdent=1&Radius=2&Radius.unit=arcmin&submit=submit+id}{SIMBAD}}\\
    4FGL J2138.9-0408 & Radio & \textcolor{blue}{\cite{2021ApJ...914...42B}}\\
    4FGL J2144.8-1600 & Radio & \textcolor{blue}{\cite{2021ApJ...914...42B}}\\
    4FGL J2212.4+0708 & Radio & \textcolor{blue}{\cite{2021ApJ...914...42B}}\\
    4FGL J2222.9+1507 & CLASS2 (unk), X-ray, Radio & \textcolor{blue}{\cite{2023arXiv230712546B}}, \textcolor{blue}{\href{http://simbad.cds.unistra.fr/simbad/sim-id?Ident=4FGL+J2222.9\%2B1507&NbIdent=1&Radius=2&Radius.unit=arcmin&submit=submit+id}{SIMBAD}}, \textcolor{blue}{\cite{2021ApJ...914...42B}}\\
    4FGL J2223.0+2704 & Radio & \textcolor{blue}{\cite{2021ApJ...914...42B}}\\
    4FGL J2225.8-0804 & X-ray, Radio & \textcolor{blue}{\cite{2021ApJ...923...75K}}, \textcolor{blue}{\cite{2021ApJ...914...42B}}\\
    4FGL J2241.4-8327 & IR & \textcolor{blue}{\href{http://simbad.cds.unistra.fr/simbad/sim-id?Ident=4FGL+J2241.4-8327&NbIdent=1&Radius=2&Radius.unit=arcmin&submit=submit+id}{SIMBAD}}\\
    4FGL J2250.5+3305 & Radio & \textcolor{blue}{\cite{2023ApJ...943...51B}}, \textcolor{blue}{\cite{2021ApJ...914...42B}}\\
    4FGL J2323.9+2519 & Radio & \textcolor{blue}{\cite{2021ApJ...914...42B}}\\
    4FGL J2348.6-2054 & IR & \textcolor{blue}{\href{http://simbad.cds.unistra.fr/simbad/sim-id?Ident=4FGL+J2348.6-2054\&NbIdent=1&Radius=5&Radius.unit=arcmin&submit=submit+id}{SIMBAD}} \\

\end{longtable}
\end{small}

\setcounter{table}{0}
\renewcommand{\thetable}{B.\arabic{table}}
\begin{small}
\setlength{\LTleft}{0pt}
\setlength{\LTright}{0pt}

\begin{center}

\begin{longtable}{%
  p{0.22\textwidth}  
  p{0.08\textwidth}  
  p{0.11\textwidth}  
  p{0.08\textwidth}  
  p{0.22\textwidth}  
  p{0.14\textwidth}  
}
\caption{Parameters of the best-fit DM models for each source in Subsample II, derived from the global likelihood fit of each model. We list the TS of the source, the preferred DM annihilation channel, the corresponding $\Delta$AIC$_c$ $=$ AIC$_c$(DM) $-$ AIC$_c$(best astrophysical model), the best-fit annihilation cross section $\langle\sigma v\rangle$ (for $J = 10^{21}$ GeV$^2$ cm$^{-5}$), and the DM particle mass $m_{\chi}$.}\\
\toprule \hline
Source Name & TS & DM Channel & $\Delta$AIC$_c$ & $\langle\sigma v\rangle$ (cm$^3$ s$^{-1}$) & $m_{\chi}$ (GeV)\\[0.5ex]
\midrule

\endfirsthead

Table B.1. \textit{(Continued)} & & & & & \\
\toprule  \hline
Source Name & TS & DM Channel & $\Delta$AIC$_c$ & $\langle\sigma v\rangle$ (cm$^3$ s$^{-1}$) & $m_{\chi}$ (GeV)\\[0.5ex]
\midrule
\endhead
4FGL J0007.4-0322 & 43.74 & $\tau^+\tau^-$ & 0.41 & $(1.429 \pm 0.652)\times 10^{-27}$ & 41.01 $\pm$ 11.24 \\
4FGL J0009.9+2810 & 44.17 & $b\bar{b}$ & -1.55 & $(3.235 \pm 0.854)\times 10^{-27}$ & 151.70 $\pm$ 29.59 \\
4FGL J0040.7-7157 & 57.34 & $W^+W^-$ & 14.13 & $(6.281 \pm 1.362)\times 10^{-27}$ & 175.20 $\pm$ 22.28 \\
4FGL J0048.6-6347 & 275.88 & $ZZ$ & 16.80 & $(3.516 \pm 0.384)\times 10^{-27}$ & 43.99 $\pm$ 3.90 \\
4FGL J0107.4+0459 & 35.51 & $\mu^+\mu^-$ & -0.45 & $(1.726 \pm 6.767)\times 10^{-26}$ & 50.63 $\pm$ 125.00 \\
4FGL J0110.0-5246 & 63.41 & $W^+W^-$ & 11.39 & $(5.208 \pm 0.898)\times 10^{-28}$ & 15.95 $\pm$ 2.29 \\
4FGL J0121.4-7216 & 32.37 & $ZZ$ & -4.53 & $(9.947 \pm 3.161)\times 10^{-27}$ & 418.60 $\pm$ 80.35 \\
4FGL J0131.7-5346 & 157.61 & $b\bar{b}$ & 14.01 & $(1.287 \pm 0.277)\times 10^{-27}$ & 31.38 $\pm$ 6.08 \\
4FGL J0140.5-4730 & 50.43 & $W^+W^-$ & 4.02 & $(4.205 \pm 0.750)\times 10^{-28}$ & 14.47 $\pm$ 1.90 \\
4FGL J0152.9-1109 & 37.70 & $ZZ$ & 0.89 & $(5.941 \pm 1.618)\times 10^{-27}$ & 224.10 $\pm$ 36.38 \\
4FGL J0213.2-8333 & 41.02 & $\tau^+\tau^-$ & -2.00 & $(2.191 \pm 1.032)\times 10^{-27}$ & 59.06 $\pm$ 16.77 \\
4FGL J0215.0-5330 & 47.21 & $\tau^+\tau^-$ & 2.47 & $(5.236 \pm 1.159)\times 10^{-28}$ & 17.76 $\pm$ 2.12 \\
4FGL J0225.5-5530 & 88.91 & $ZZ$ & 1.90 & $(7.117 \pm 1.193)\times 10^{-27}$ & 132.90 $\pm$ 14.77 \\
4FGL J0243.1-4218 & 43.43 & $ZZ$ & -1.00 & $(1.593 \pm 0.566)\times 10^{-27}$ & 70.70 $\pm$ 23.12 \\
4FGL J0259.1-4748 & 51.46 & $\mu^+\mu^-$ & 15.11 & $(4.789 \pm 1.021)\times 10^{-27}$ & 13.89 $\pm$ 0.69 \\
4FGL J0309.9-6941 & 34.18 & $e^+e^-$ & 14.93 & $(8.292 \pm 1.976)\times 10^{-28}$ & 5.53 $\pm$ 0.97 \\
4FGL J0347.0-6400 & 59.23 & $\tau^+\tau^-$ & 4.17 & $(2.791 \pm 0.459)\times 10^{-28}$ & 6.13 $\pm$ 0.51 \\
4FGL J0357.7-6808 & 47.23 & $ZZ$ & 20.31 & $(7.813 \pm 1.736)\times 10^{-27}$ & 215.70 $\pm$ 27.25 \\
4FGL J0413.4-8609 & 76.87 & $b\bar{b}$ & 0.73 & $(1.344 \pm 0.211)\times 10^{-27}$ & 35.97 $\pm$ 4.58 \\
4FGL J0423.0-0943 & 118.82 & $b\bar{b}$ & 0.05 & $(1.671 \pm 0.348)\times 10^{-27}$ & 34.93 $\pm$ 6.38 \\
4FGL J0506.6-4629 & 86.68 & $ZZ$ & 7.21 & $(9.800 \pm 3.977)\times 10^{-27}$ & 239.50 $\pm$ 66.17 \\
4FGL J0515.4-3841 & 44.57 & $ZZ$ & 2.25 & $(4.566 \pm 1.079)\times 10^{-27}$ & 171.80 $\pm$ 23.81 \\
4FGL J0558.1-2113 & 166.09 & $\mu^+\mu^-$ & -2.83 & $(5.510 \pm 0.656)\times 10^{-27}$ & 5.53 $\pm$ 0.31 \\
4FGL J0712.0-6431 & 124.49 & $\tau^+\tau^-$ & 2.17 & $(7.410 \pm 0.948)\times 10^{-28}$ & 9.91 $\pm$ 0.78 \\
4FGL J0744.0+5033 & 44.69 & $\mu^+\mu^-$ & 48.07 & $(9.606 \pm 2.309)\times 10^{-27}$ & 24.47 $\pm$ 2.98 \\
4FGL J0944.3-0911 & 37.31 & $ZZ$ & -0.15 & $(2.839 \pm 0.954)\times 10^{-27}$ & 98.31 $\pm$ 23.90 \\
4FGL J1022.1-1818 & 47.06 & $c\bar{c}$ & -1.37 & $(3.366 \pm 0.948)\times 10^{-27}$ & 136.90 $\pm$ 28.31 \\
4FGL J1044.5-1439 & 50.93 & $\mu^+\mu^-$ & -1.45 & $(2.233 \pm 9.544)\times 10^{-27}$ & 23.94 $\pm$ 7.79 \\
4FGL J1105.1-2600 & 125.43 & $b\bar{b}$ & -1.82 & $(1.658 \pm 0.305)\times 10^{-27}$ & 37.27 $\pm$ 5.46 \\
4FGL J1105.3-0626 & 37.67 & $ZZ$ & 2.66 & $(2.123 \pm 0.820)\times 10^{-26}$ & 854.10 $\pm$ 232.60 \\
4FGL J1213.6+1236 & 37.25 & $\mu^+\mu^-$ & 3.85 & $(6.159 \pm 1.704)\times 10^{-27}$ & 19.37 $\pm$ 2.90 \\
4FGL J1219.8-2008 & 47.31 & $e^+e^-$ & 6.44 & $(5.370 \pm 1.325)\times 10^{-27}$ & 24.97 $\pm$ 2.99 \\
4FGL J1225.9+2951 & 1687.47 & $b\bar{b}$ & -8.16 & $(1.222 \pm 0.127)\times 10^{-26}$ & 45.10 $\pm$ 3.77 \\
4FGL J1400.0-2415 & 500.81 & $c\bar{c}$ & 0.17 & $(3.288 \pm 0.295)\times 10^{-27}$ & 30.79 $\pm$ 2.25 \\
4FGL J1405.9-1853 & 51.11 & $e^+e^-$ & -3.98 & $(7.548 \pm 1.939)\times 10^{-27}$ & 32.31 $\pm$ 3.51 \\
4FGL J1420.9-2431 & 179.16 & $ZZ$ & -1.73 & $(4.352 \pm 0.479)\times 10^{-27}$ & 49.33 $\pm$ 3.93 \\
4FGL J1432.1+5020 & 47.53 & $W^+W^-$ & 0.07 & $(2.173 \pm 0.490)\times 10^{-27}$ & 96.75 $\pm$ 15.11 \\
4FGL J1526.9+7358 & 26.41 & $\mu^+\mu^-$ & 38.39 & $(2.496 \pm 0.651)\times 10^{-27}$ & 10.33 $\pm$ 0.37 \\
4FGL J1528.2-2448 & 44.27 & $c\bar{c}$ & 0.49 & $(1.091 \pm 0.201)\times 10^{-27}$ & 33.30 $\pm$ 4.52 \\
4FGL J1547.0-0442 & 64.78 & $ZZ$ & 2.18 & $(6.254 \pm 4.601)\times 10^{-26}$ & 862.70 $\pm$ 451.60 \\
4FGL J1612.4+0409 & 123.78 & $b\bar{b}$ & -0.69 & $(3.159 \pm 0.453)\times 10^{-27}$ & 59.79 $\pm$ 6.82 \\
4FGL J1630.1-1049 & 203.47 & $\tau^+\tau^-$ & 1.81 & $(1.527 \pm 0.171)\times 10^{-27}$ & 9.05 $\pm$ 0.70 \\
4FGL J1635.3+4258 & 179.73 & $\tau^+\tau^-$ & 0.03 & $(7.879 \pm 1.228)\times 10^{-28}$ & 12.29 $\pm$ 1.27 \\
4FGL J1655.6+0317 & 34.04 & $e^+e^-$ & -0.87 & $(1.121 \pm 1.111)\times 10^{-26}$ & 44.91 $\pm$ 9108.00 \\
4FGL J1656.4-0410 & 106.12 & $b\bar{b}$ & 1.73 & $(2.549 \pm 0.528)\times 10^{-27}$ & 35.30 $\pm$ 6.72 \\
4FGL J1659.0-0140 & 315.49 & $ZZ$ & 0.20 & $(8.706 \pm 0.746)\times 10^{-27}$ & 47.53 $\pm$ 2.91 \\
4FGL J1809.0+5908 & 38.99 & $W^+W^-$ & 20.61 & $(2.551 \pm 0.718)\times 10^{-27}$ & 142.70 $\pm$ 309.60 \\
4FGL J1816.4-6405 & 246.55 & $ZZ$ & -1.30 & $(2.021 \pm 0.183)\times 10^{-27}$ & 15.70 $\pm$ 1.18 \\
4FGL J1827.2+4303 & 61.47 & $e^+e^-$ & -2.18 & $(2.292 \pm 0.425)\times 10^{-27}$ & 10.38 $\pm$ 0.75 \\
4FGL J1905.2-5120 & 75.56 & $W^+W^-$ & 1.85 & $(8.794 \pm 1.253)\times 10^{-28}$ & 11.35 $\pm$ 0.58 \\
4FGL J1956.6-7011 & 95.02 & $e^+e^-$ & 2.65 & $(1.902 \pm 0.284)\times 10^{-27}$ & 5.64 $\pm$ 0.37 \\
4FGL J2007.2-3157 & 109.69 & $c\bar{c}$ & -1.09 & $(2.420 \pm 0.321)\times 10^{-27}$ & 44.54 $\pm$ 3.96 \\
4FGL J2043.9-4802 & 360.07 & $c\bar{c}$ & 10.05 & $(2.540 \pm 0.213)\times 10^{-27}$ & 27.85 $\pm$ 1.75 \\
4FGL J2055.0-5218 & 35.37 & $e^+e^-$ & 29.48 & $(1.058 \pm 0.234)\times 10^{-27}$ & 5.94 $\pm$ 0.67 \\
4FGL J2057.6-7829 & 48.43 & $e^+e^-$ & 49.47 & $(2.296 \pm 0.636)\times 10^{-27}$ & 12.11 $\pm$ 1.12 \\
4FGL J2122.5+0345 & 57.13 & $c\bar{c}$ & 15.06 & $(2.881 \pm 0.387)\times 10^{-28}$ & 6.00 $\pm$ 0.47 \\
4FGL J2133.1-6432 & 689.49 & $ZZ$ & 0.55 & $(2.864 \pm 0.208)\times 10^{-27}$ & 13.52 $\pm$ 0.85 \\
4FGL J2151.2-4034 & 37.21 & $e^+e^-$ & 20.66 & $(3.458 \pm 0.846)\times 10^{-27}$ & 13.74 $\pm$ 2.10 \\
4FGL J2207.3+2746 & 28.70 & $e^+e^-$ & 8.03 & $(1.641 \pm 0.446)\times 10^{-27}$ & 11.01 $\pm$ 1.38 \\
4FGL J2217.0-6727 & 55.33 & $b\bar{b}$ & 7.27 & $(1.194 \pm 0.627)\times 10^{-26}$ & 417.50 $\pm$ 150.40 \\
4FGL J2219.4+3229 & 46.56 & $ZZ$ & 2.77 & $(9.471 \pm 3.056)\times 10^{-27}$ & 259.80 $\pm$ 63.17 \\
4FGL J2318.6-3827 & 84.67 & $W^+W^-$ & 4.51 & $(1.525 \pm 0.263)\times 10^{-27}$ & 37.38 $\pm$ 4.93 \\
4FGL J2342.4-4739 & 342.93 & $c\bar{c}$ & -9.57 & $(1.574 \pm 0.145)\times 10^{-27}$ & 23.75 $\pm$ 1.64 \\
4FGL J2343.0-4756 & 66.50 & $\tau^+\tau^-$ & 0.04 & $(1.212 \pm 0.251)\times 10^{-27}$ & 25.82 $\pm$ 2.73 \\
\hline
\bottomrule
\label{table-dmfit-aicc}
\end{longtable}
\end{center}
\end{small}

\end{document}